# A Total-Variation Sparseness-Promoting Method for the Synthesis of Contiguously Clustered Linear Arrays

Nicola Anselmi, *Member, IEEE*, Giorgio Gottardi, *Member, IEEE*, Giacomo Oliveri, *Senior Member, IEEE*, and Andrea Massa, *Fellow, IEEE*

***Abstract*—By exploiting an innovative Total-Variation Compressive Sensing (*TV-CS*) formulation, a new method for the synthesis of physically-contiguous clustered linear arrays is presented. The computation of the feed network excitations is recast as the maximization of the *gradient sparsity* of the excitation vector subject to matching a user-defined pattern. The arising *TV-CS* functional is then optimized by means of a deterministic *alternating direction* algorithm. A selected set of representative numerical results, drawn from a wide validation, is reported to illustrate the potentialities and the limitations of the proposed approach when clustering arrays of both ideal and realistic antenna elements. Comparisons with some competitive state-of-the-art subarraying techniques are performed, as well.**

## I. Introduction and Rationale

THE CAREFUL control of the beam shape [1][2][3] enabled by an antenna array is a key feature in those applicative scenarios, such as remote sensing [4], radar and radiolocalization [5], wireless power transmission [6][7][8], and satellite [9][10][11] and *5G* [12] communications, where the radiated pattern and the sidelobes must fit arbitrary user-defined masks. Although *fully-populated* (*FP*) arrays (i.e., conventional architectures comprising one control point for each radiating element) are effective solutions in terms of radiation performances, often they are too expensive to deploy because of the power consumption and the costs for the fabrication/testing/maintenance of the corresponding feed networks [2][13]. Therefore, unconventional designs aimed at reducing the circuitry costs and complexity, while exhibiting suitable beam-shaping capabilities and an acceptable control of the radiated pattern, have been introduced [13] as competitive alternatives to *FP* arrays.

In such a framework, unconventional clustered layouts [12][13][14], which are $N$-elements standard arrays driven by only $Q$ ($Q < N$) control points, have been studied in several works [12][13][14][15][16][17]. Because of the dimensionality of the problem at hand, the arising synthesis process turns out very challenging from both the theoretical and the practical viewpoint [15][16][17][18] even more if the physical contiguity of the clusters [i.e., the fact that each control point can drive only contiguous radiators, or in other words, that all elements belonging to a cluster are physically adjacent - Fig. 1(*a*)] is required to minimize the feed network complexity and the electromagnetic coupling effects [13]. Since theoretically-optimal subarraying methods - introduced in [19] and later on hybridized in [20] - do not *a-priori* guarantee the contiguity-constraint [19][20], sub-optimal techniques have

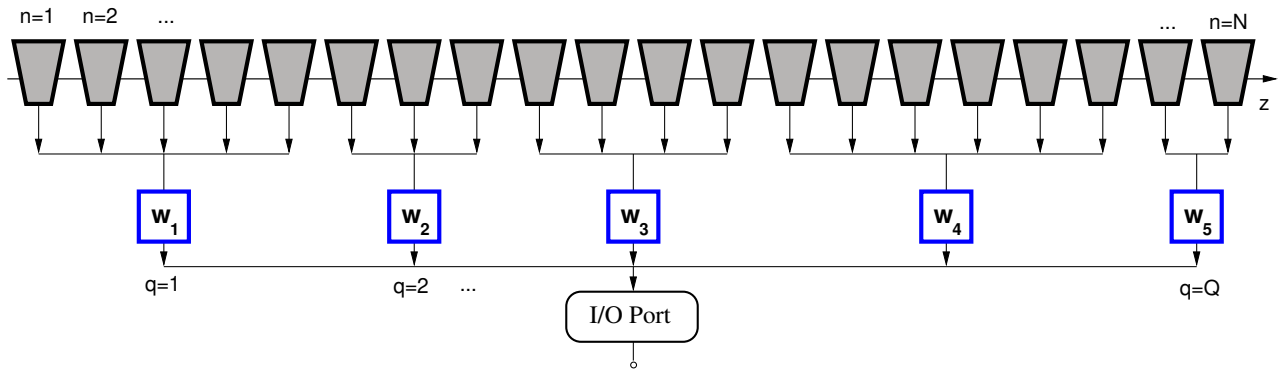

(*a*)

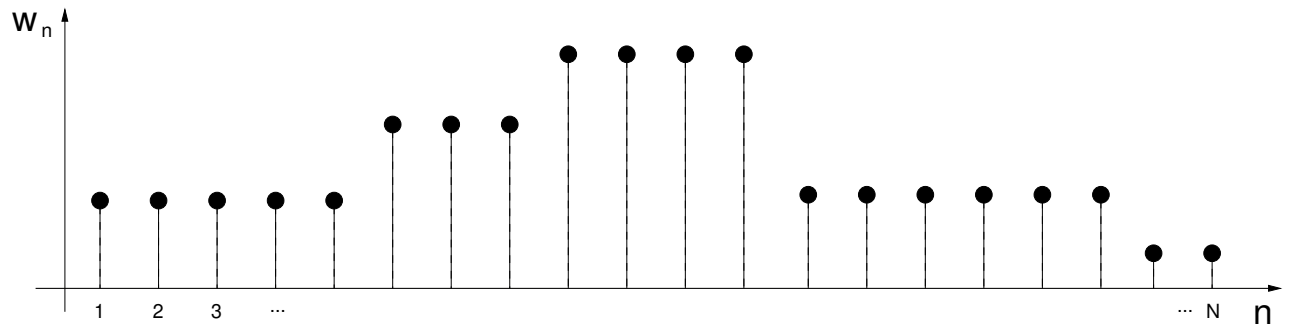

(*b*)

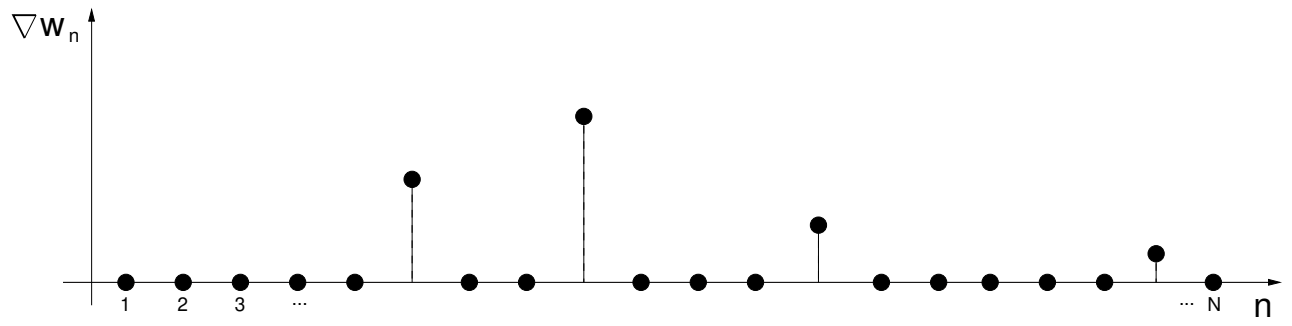

(*c*)

Figure 1. *Problem Formulation* - Example of (*a*) a clustered layout, (*b*) the corresponding excitation coefficients $\mathbf{w}$, and (*c*) the associated discrete gradient $\nabla\mathbf{w}$.

Manuscript received May XX, 2018

This work benefited from the networking activities carried out within the Project Antenne al Plasma - Tecnologia abilitante per SATCOM (ASI.EPT.COM) funded by the Italian Space Agency (ASI) under Grant 2018-3-HH.0 (CUP: F91I17000020005), the SNATCH Project (2017-2019) founded by the Italian Ministry of Foreign Affairs and International Cooperation, Directorate General for Cultural and Economic Promotion and Innovation, and the Project "CLOAKING METASURFACES FOR A NEW GENERATION OF INTELLIGENT ANTENNA SYSTEMS (MANTLES)" funded by the Italian Ministry of Education, University, and Research within the PRIN2017 Program.

N. Anselmi, G. Gottardi, G. Oliveri and A. Massa are with the ELEDIA Research Center (ELEDIA@UniTN - University of Trento), Via Sommarive 9, 38123 Trento - Italy (e-mail: {nicola.anselmi.1, giorgio.gottardi, giacomo.oliveri, andrea.massa}@unitn.it).

G. Oliveri and A. Massa are also with the ELEDIA Research Center (ELEDIA@L2S - UMR 8506), 3 rue Joliot Curie, 91192 Gif-sur-Yvette - France (e-mail: giacomo.oliveri, andrea.massa@l2s.centralesupelec.fr).

A. Massa is also with the ELEDIA Research Center (ELEDIA@UESTC - University of Electronic Science and Technology of China), School of Electronic Engineering, 611731 Chengdu - China (e-mail: andrea.massa@uestc.edu.cn).

been developed to synthesize architectures featuring contiguous sub-arrays [16][21][22] starting from random strategies [16] and hybrid techniques [21][23][24] based on evolutionary algorithms [25][26] until *ad-hoc* versions of the Contiguous Partition Method (*CPM*) [22].

More recently, a numerically-efficient Compressive Sensing (*CS*)-based approach has been presented in [14] to *a-priori* enforce the physical contiguity of the clustered elements by specifying the *dictionary* of the candidate sub-array configurations. Thanks to the effectiveness of the *CS* paradigm [27][28] in dealing with sparse problems, the arising sub-arrayed layouts have shown excellent pattern control capabilities [14], but without avoiding some overlaps between adjacent clusters [14]. Of course, this latter (potential) drawback can be overcome by limiting the set of *dictionary* configurations so that overlapping is *a-priori* avoided, but this countermeasure would correspond to fix (in advance) the cluster lengths and positions, thus greatly reducing the degrees-of-freedom (*DoF*s) available in the design process.

Alternatively and taking inspiration by the recent theoretical advances on *CS* as applied to microwave imaging [29], a reformulation of the synthesis problem in a "transformed domain" may be profitably considered. Towards this end, let us observe that the distribution of the excitation coefficients of a contiguously-clustered layout turns out to be a *piecewise constant* function [Fig. 1(*b*)], then the *discrete gradient* of the excitation vector is naturally *sparse* since its entries are non-null only in correspondence with the "border elements" between adjacent clusters [Fig. 1(*c*)]. Therefore, a *CS*-based contiguous clustering strategy can be formulated by enforcing the sparsity prior on the *discrete gradient* of the array excitations [i.e., to minimize the number of sub-arrays in the feed network - Fig. 1(*c*) vs. Fig. 1(*a*)] rather than on the combination of clusters from a user-defined dictionary such as in [14]. By adapting to the contiguous clustering problem the Total-Variation *CS* (*TV-CS*) formulation (a *CS* approach intrinsically encoding the sparsity of the *discrete gradient* of the unknown as a prior [29]), it would be then possible (*i*) to fully exploit the effectiveness and the efficiency of the *CS* paradigm; (*ii*) to *a-priori* avoid the sub-arrays overlapping; (*iii*) to avoid the *a-priori* setup of the positions and the lengths of the array clusters, thus enabling the full exploration of the design *DoF*s; (*iv*) to minimize by definition the costs/complexity of the feeding architecture (i.e., the number of control points) since minimizing the gradient sparsity is equivalent to reduce the number of clusters [Figs. 1(*b*)-1(*c*)]; (*v*) to potentially customize to the clustering problem at hand several effective and numerically efficient solvers that have been developed in the state-of-the-art literature for *TV-CS* inversions [30][31][32][33][34][35][36].

Accordingly, an innovative method based on the *TV-CS* formulation for the synthesis of contiguously clustered linear arrays is proposed in this work. More specifically, the design of the optimally sub-arrayed element excitations is carried out by maximizing the sparsity of the antenna excitations in the *gradient* domain. Because of the non-differentiability of the arising functional, the *TV-CS* formulation is firstly recast as the minimization of an Augmented Lagrangian (*AL*) one, then a deterministic *alternating direction* algorithm [30][31][32] is applied for its solution. Such a choice is suggested by the advantages in terms of convergence rate when handling a large number of *DoF*s [33][34][35] and of flexibility (i.e., it does not require restrictive assumptions on the observation matrix, unlike [36]) of the resulting solution scheme [30][31][32] with respect to other solvers (e.g., *SOCP* [33], "$\ell_1$-Magic" [34], *TwIST* [35], and *NESTA* [36]) already adopted in the *TV-CS* framework.

As for the innovative methodological contributions of this work, the most relevant ones include (*i*) the formulation of the clustering problem as a sparseness-promoting one that, unlike previous *CS*-based techniques [14], operates in the *gradient* domain and it is not constrained to *a-priori* define a *cluster-shape dictionary*, (*ii*) the derivation of a *CS* sub-arraying algorithm that by definition avoids overlapping without *a-priori* limiting the *DoF*s, (*iii*) the exploitation and the customization to array synthesis of the *AL*-based technique proposed in [30][31][32], and (*iv*) the direct formulation in the complex-excitations domain of the *CS*-based array clustering strategy, thus avoiding incoherent clustered arrangements (i.e., different arrangements for the real and the imaginary part of the excitation coefficients), unlike many existing *CS* design methods [27][37] that deal separately with the real and the imaginary part of the weights.

The outline of the paper is as follows. After the formulation of the physically-contiguous clustering problem in the *CS* framework (Sect. II), the *TV-CS* based synthesis strategy is described (Sect. III). Selected numerical examples are then presented to give the interested readers some insights on the effectiveness and the potentialities of the proposed clustering method also in comparison with state-of-the-art techniques (Sect. IV). Finally, some conclusions and final remarks are drawn in Sect. V.

## II. Statement of the Pattern-Matching Contiguous-Clustering *CS* Problem

Let us consider the problem of determining the set of excitation coefficients, $\mathbf{w} \triangleq \{w_n, n = 1, ..., N\}$, of a linear clustered array of $N$ elements [Fig. 1(*a*)] displaced along the $\widehat{z}$-axis at the locations $z_n$, $n = 1, ..., N$, so that it radiates a far field[1] pattern [1][2]

$$f(\theta) = \sum_{n=1}^{n} w_n e_n(\theta) \exp\left[j\frac{2\pi z_n \sin(\theta)}{\lambda}\right], \qquad (1)$$

that matches a desired one, $\widetilde{f}(\theta)$, $\lambda$ and $e_n(\theta)$ being the free-space wavelength and the embedded element pattern of the $n$-th array element. In other words, the *pattern-matching contiguous-clustering problem* at hand is that of finding the *clustered* excitations starting from the knowledge of the array geometry (i.e., $z_n$, $n = 1, .., , N$) and of $\widetilde{f}(\theta)$. Such a formulation in the *excitation-domain* [37][38] does not comply with the applicability conditions of the standard *CS*, the unknown

[1]For the sake of notation simplicity, the radial dependence (i.e., $\frac{\exp(-jkr)}{r}$ term) is assumed and omitted hereinafter and only the co-polar component of the field is considered. Of course, the method and its formulation can be straightforwardly extended to the vectorial case.

vector $\mathbf{w}$ being not sparse since it mathematically 'codes' a *FP* arrangement [Fig. 1(*b*)]. However, the sparsity condition can be recovered in the *gradient-excitation domain* by observing that a clustered arrangement exhibits a piecewise-constant $\mathbf{w}$ vector [Fig. 1(*b*)], then its gradient, $\nabla\mathbf{w}$, turns out to be sparse having few non-zero entries located only in correspondence with the "border elements" between physically-contiguous sub-arrays [Fig. 1(*c*)]. Accordingly, our clustering problem can profitably recast in the *CS* framework by exploiting the following *TV-CS* formulation [29][30][31][32]

$$\min_{\mathbf{w}} \|\nabla\mathbf{w}\|_1 \text{ s.t. } f(\theta) = \widetilde{f}(\theta) \tag{2}$$

where $\|\cdot\|_1$ stands for the $\ell_1$-norm. The essence of the efficacy of the *TV-CS* in antenna array synthesis is actually motivated by the fact that (2) explicitly aims at the minimization of the number of clusters in the final architecture [i.e., $\min_{\mathbf{w}} \|\nabla\mathbf{w}\|_1$ objective - (2)] subject to a pattern matching requirement [i.e., $f(\theta) = \widetilde{f}(\theta)$ constraint - (2)], thus naturally encoding the designer's need to reduce the feed network costs and complexity while exhibiting a suitable beam-shape control. In matrix form, (2) reads as follows [29][30][31][32]

$$\min_{\boldsymbol{\alpha},\mathbf{w}} \|\boldsymbol{\alpha}\|_1 \text{ s.t. } \boldsymbol{\alpha} = \nabla\mathbf{w} \text{ and } \mathcal{H}\mathbf{w} = \widetilde{\mathbf{f}} \tag{3}$$

where

$$\mathcal{H} \triangleq \left\{h_{mn} = e_n(\theta_m)\exp\left[j\frac{2\pi z_n \sin(\theta_m)}{\lambda}\right],\; m = 1,...,M,\, n = 1,...,N\right\} \tag{4}$$

is the discretized array radiation operator, $\widetilde{\mathbf{f}} \triangleq \left\{\widetilde{f}(\theta_m),\, m = 1,...,M\right\}$ is the vector whose $m$-th entry is the sample of the target field at the user-defined direction $\theta_m$ ($m = 1,...,M$). Thanks to the introduction of the auxiliary vector $\boldsymbol{\alpha} \triangleq \{\alpha_n,\, n = 1,...,N\}$ [29][30][31][32] and by using the Lagrangian multiplier method [39], (3) can be re-written into the following *convex* equivalent expression

$$\min_{\boldsymbol{\alpha},\mathbf{w},\boldsymbol{\mu},\boldsymbol{\eta}} \left[\Phi_{\mathcal{L}}(\boldsymbol{\alpha},\mathbf{w},\boldsymbol{\mu},\boldsymbol{\eta})\right] \text{ s.t. } \boldsymbol{\alpha} = \nabla\mathbf{w} \text{ and } \mathcal{H}\mathbf{w} = \widetilde{\mathbf{f}} \tag{5}$$

where $\boldsymbol{\mu} \triangleq \{\mu_n,\, n = 1,...,N\}$ and $\boldsymbol{\eta} \triangleq \{\eta_n,\, n = 1,...,N\}$ are the Lagrangian multiplier vectors, $\Phi_{\mathcal{L}}(\boldsymbol{\alpha},\mathbf{w},\boldsymbol{\mu},\boldsymbol{\eta})$ is the Lagrangian function

$$\Phi_{\mathcal{L}}(\boldsymbol{\alpha},\mathbf{w},\boldsymbol{\mu},\boldsymbol{\eta}) \triangleq \|\boldsymbol{\alpha}\|_1 - \boldsymbol{\mu}^*(\nabla\mathbf{w} - \boldsymbol{\alpha}) - \boldsymbol{\eta}^*\left(\mathcal{H}\mathbf{w} - \widetilde{\mathbf{f}}\right), \tag{6}$$

and $\cdot^*$ is the conjugate transpose operator [29][39].

It is worth pointing out that, unlike [14], such a *CS* formulation of the *pattern-matching contiguous-clustering problem* (*i*) does not require the user to *a-priori* specify a *dictionary* of candidate sub-arrays, and (*ii*) it guarantees - by definition - no overlap among the array elements belonging to different sub-arrays.

## III. *TV* Sparseness-Promoting Solution Strategy

Despite its convexity, solving the *TV* clustering problem stated in Sect. II is still not a trivial task because of the non-linear, the non-differentiable, and the constrained nature of (5). To efficiently and effectively address such challenges, the use of the penalty term $\Phi_{\mathcal{P}}(\boldsymbol{\alpha},\mathbf{w},\boldsymbol{\mu},\boldsymbol{\eta})$ is firstly considered to yield the following *unconstrained* cost function minimization problem

$$\min_{\boldsymbol{\alpha},\mathbf{w},\boldsymbol{\mu},\boldsymbol{\eta}} \left[\Phi(\boldsymbol{\alpha},\mathbf{w},\boldsymbol{\mu},\boldsymbol{\eta})\right] \tag{7}$$

where $\Phi(\boldsymbol{\alpha},\mathbf{w},\boldsymbol{\mu},\boldsymbol{\eta})$ is the Augmented Lagrangian cost function

$$\Phi(\boldsymbol{\alpha},\mathbf{w},\boldsymbol{\mu},\boldsymbol{\eta}) \triangleq \Phi_{\mathcal{L}}(\boldsymbol{\alpha},\mathbf{w},\boldsymbol{\mu},\boldsymbol{\eta}) + \Phi_{\mathcal{P}}(\boldsymbol{\alpha},\mathbf{w},\boldsymbol{\mu},\boldsymbol{\eta}), \tag{8}$$

and $\Phi_{\mathcal{P}}(\boldsymbol{\alpha},\mathbf{w},\boldsymbol{\mu},\boldsymbol{\eta})$ is the weighted *mismatch* function

$$\Phi_{\mathcal{P}}(\boldsymbol{\alpha},\mathbf{w},\boldsymbol{\mu},\boldsymbol{\eta}) \triangleq \frac{\beta}{2}\|\nabla\mathbf{w} - \boldsymbol{\alpha}\|_2^2 + \frac{\gamma}{2}\left\|\mathcal{H}\mathbf{w} - \widetilde{\mathbf{f}}\right\|_2^2, \tag{9}$$

that "models" the constraints in (5), $\|\cdot\|_2$ being the $\ell_2$-norm, while $\beta$ and $\gamma$ are the user-defined *barrier* parameters.

Under such assumptions and according to the Lagrangian approach, the optimal values of the multipliers, $\widehat{\boldsymbol{\mu}} = \boldsymbol{\mu}^{(K)}$ and $\widehat{\boldsymbol{\eta}} = \boldsymbol{\eta}^{(K)}$, can be found by an iterative ($k$ being the iteration index, $k = 1,...,K$) search procedure based on the following update formulas [30]

$$\boldsymbol{\mu}^{(k+1)} = \boldsymbol{\mu}^{(k)} - \beta\left(\nabla\mathbf{w}^* - \boldsymbol{\alpha}^*\right) \tag{10}$$

$$\boldsymbol{\eta}^{(k+1)} = \boldsymbol{\eta}^{(k)} - \gamma\left(\mathcal{H}\mathbf{w}^* - \widetilde{\mathbf{f}}^*\right) \tag{11}$$

once determined the unknown vectors $\boldsymbol{\alpha}$ and $\mathbf{w}$. Towards this end, let us observe that the Augmented Lagrangian cost function $\Phi$ in (7) is non-differentiable owing to the presence of the $\|\cdot\|_1$-term in $\Phi_{\mathcal{L}}$, thus it cannot be addressed through a standard (local) search procedure. However, by analyzing the functional dependence of (8) on the unknown vectors $\boldsymbol{\alpha}$ and $\mathbf{w}$, it turns out that

- solving (8) with respect to $\boldsymbol{\alpha}$ gives the closed-form solution [30][31]

$$\widehat{\boldsymbol{\alpha}}(\mathbf{w},\boldsymbol{\mu},\boldsymbol{\eta}) = \left\{\widehat{\alpha}_n = \max\left\{\left|\nabla\mathbf{w}|_n - \frac{\mu_n}{\beta}\right| - \frac{1}{\beta}, 0\right\} \times \operatorname{sgn}\left(\nabla\mathbf{w}|_n - \frac{\mu_n}{\beta}\right),\, n = 1,...,N\right\} \tag{12}$$

  where $\operatorname{sgn}(\cdot)$ is the sign operator and $\nabla\mathbf{w}|_n$ is the $n$-th entry of $\nabla\mathbf{w}$;
- $\Phi(\boldsymbol{\alpha},\mathbf{w},\boldsymbol{\mu},\boldsymbol{\eta})$ has a quadratic dependence on $\mathbf{w}$ and its gradient can be computed as follows [30][31]

$$\mathbf{d}(\boldsymbol{\alpha},\mathbf{w},\boldsymbol{\mu},\boldsymbol{\eta}) = \nabla^*(-\beta\nabla\mathbf{w} - \beta\boldsymbol{\alpha} - \boldsymbol{\mu}) + {} + \gamma\mathcal{H}^*\left(\mathcal{H}\mathbf{w} - \widetilde{\mathbf{f}} - \boldsymbol{\eta}\right). \tag{13}$$

Taking advantage from these features, an efficient approximated methodology that minimizes $\Phi(\boldsymbol{\alpha},\mathbf{w},\boldsymbol{\mu},\boldsymbol{\eta})$ by using (10)-(11) and updating $\boldsymbol{\alpha}$ and $\mathbf{w}$ in an *alternate* fashion can be profitably exploited to address (7). More specifically, starting from (*a*) an initial ($k = 1$) configuration of the multipliers and array weights, $\boldsymbol{\mu}^{(1)} = \boldsymbol{\eta}^{(1)} = \mathbf{0}$, $\mathbf{w}^{(1)} = \mathbf{0}$), (*b*) the value of $\boldsymbol{\alpha}^{(k+1)}$ is yielded as $\boldsymbol{\alpha}^{(k+1)} = \widehat{\boldsymbol{\alpha}}\left(\mathbf{w}^{(k)},\boldsymbol{\mu}^{(k)},\boldsymbol{\eta}^{(k)}\right)$ according to (12), then (*c*) the excitations are updated by adopting a *one-step steepest descent method* [29][30]

$$\mathbf{w}^{(k+1)} = \mathbf{w}^{(k)} - \rho^{(k)}\sigma^{(k)}\mathbf{d}^{(k)}, \tag{14}$$

where $\mathbf{d}^{(k)} \triangleq \mathbf{d}\left(\boldsymbol{\alpha}^{(k)},\mathbf{w}^{(k)},\boldsymbol{\mu}^{(k)},\boldsymbol{\eta}^{(k)}\right)$ (13), $\sigma^{(k)}$ is the $k$-th "aggressive" step length, which is computed through the Barzilai-Borwein formula

$$\sigma^{(k)} = \frac{\left(\mathbf{w}^{(k)} - \mathbf{w}^{(k-1)}\right)^*\left(\mathbf{w}^{(k)} - \mathbf{w}^{(k-1)}\right)}{\left(\mathbf{w}^{(k)} - \mathbf{w}^{(k-1)}\right)^*\left(\mathbf{d}^{(k)} - \mathbf{d}^{(k-1)}\right)}, \tag{15}$$

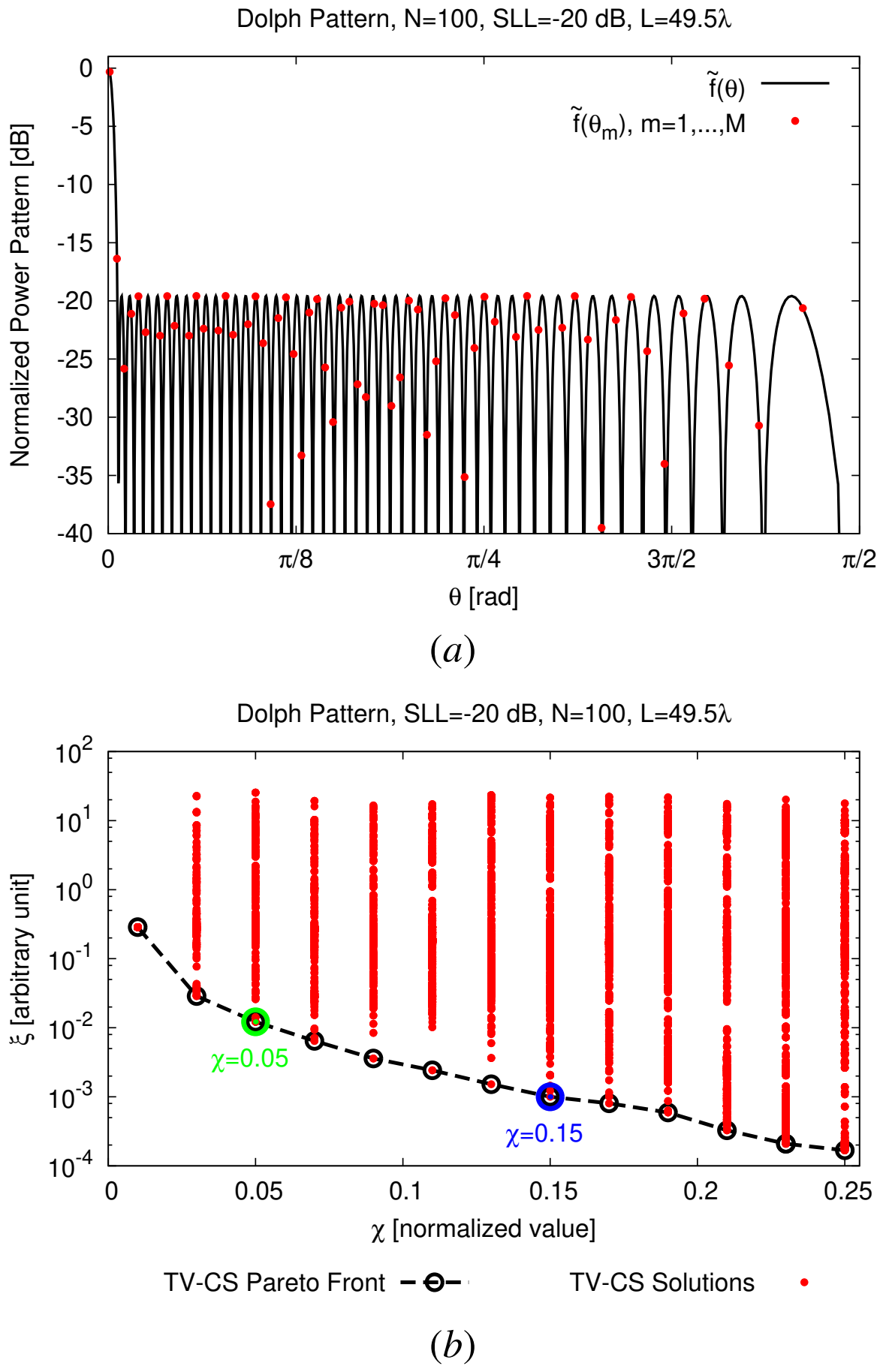

Figure 2. *Illustrative Example* ($N=100$, *Dolph Pattern* [$SLL=-20$ *dB*]) - (*a*) Plot of $\widetilde{f}(\theta)$ and $\widetilde{f}(\theta_m)$, $m=1,...,M$, and (*b*) representative *TV-CS* solutions and associated Pareto fronts in the $(\xi,\chi)$-plane.

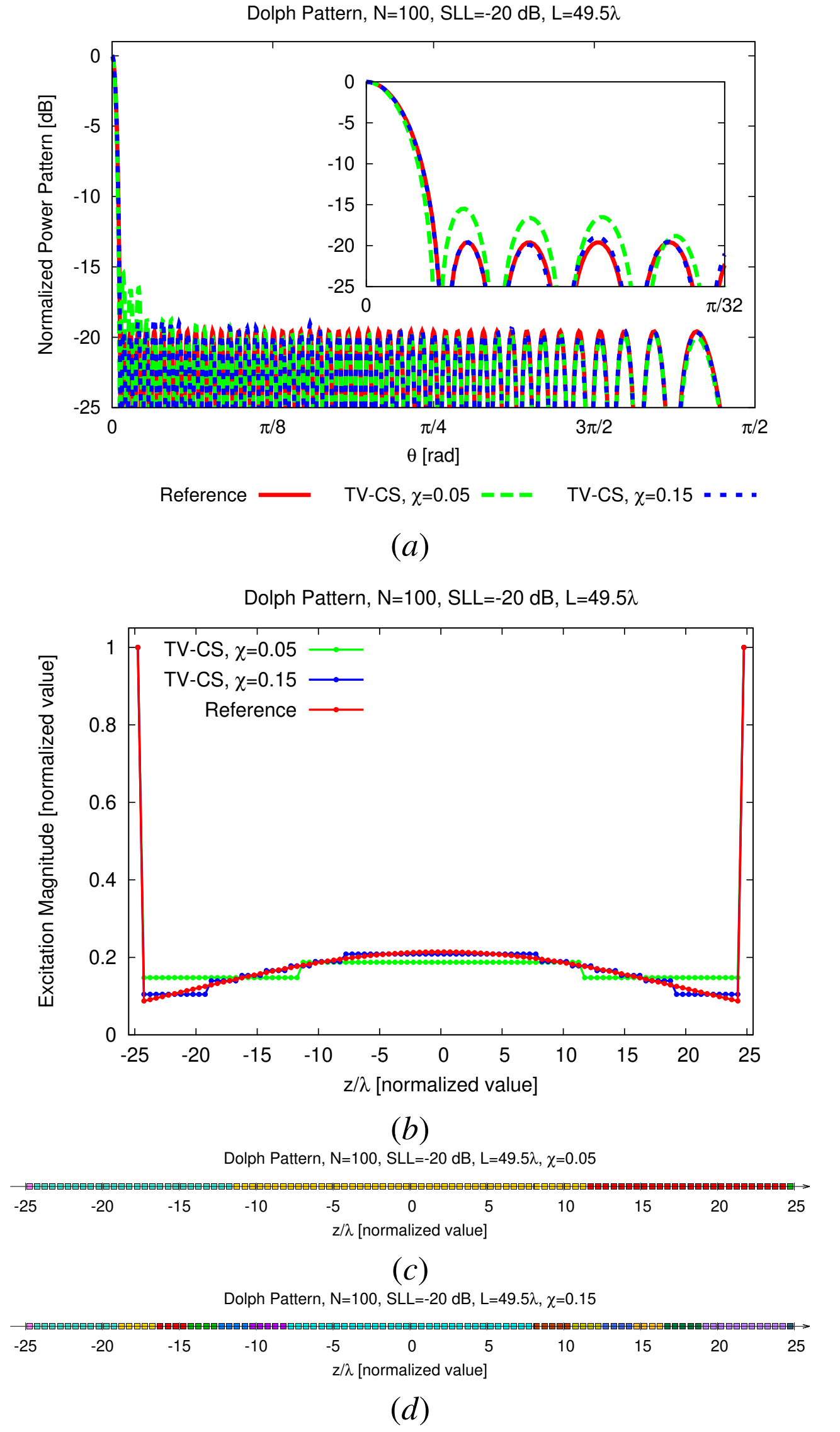

Figure 3. *Numerical Validation* ($N=100$, *Dolph Pattern* [$SLL=-20$ *dB*]) - (*a*) Reference and *TV-CS* patterns, (*b*) corresponding excitations, and (*c*)(*d*) clustering schemes (each color indicates the membership to a cluster) when (*a*)(*b*)(*c*) $\chi=0.05$ and (*a*)(*b*)(*d*) $\chi=0.15$.

and $\rho^{(k)}\in[0,1]$ is the $k$-th iteration scaling coefficient chosen so that the Armijo condition

$$\frac{\Phi\left(\boldsymbol{\alpha}^{(k)},\mathbf{w}^{(k)},\boldsymbol{\mu}^{(k)},\boldsymbol{\eta}^{(k)}\right)-\Phi\left(\boldsymbol{\alpha}^{(k)},\mathbf{w}^{(k+1)},\boldsymbol{\mu}^{(k)},\boldsymbol{\eta}^{(k)}\right)}{\rho^{(k)}\sigma^{(k)}\left(\mathbf{d}^{(k)}\right)^{*}\mathbf{d}^{(k)}}\geq\nu \tag{16}$$

holds true, $\nu$ being a user-defined parameter. Finally, (*d*) the multipliers are updated according to (10)-(11). The steps (*b*)-(*c*)-(*d*) are then iterated until either $k=K$ or when fitting the following convergence condition

$$\frac{\left\|\mathbf{w}^{(k+1)}-\mathbf{w}^{(k)}\right\|_2}{\left\|\mathbf{w}^{(k)}\right\|_2}\leq\delta\quad\text{if } k=\widehat{K} \tag{17}$$

where $\delta$ is the user-defined convergence threshold, to finally yield the optimized excitations

$$\widehat{\mathbf{w}}=\mathbf{w}^{\left(\min\left\{K,\widehat{K}\right\}\right)}.$$

It is worthwhile to notice that the control parameters of such a *TV* Sparseness-Promoting Solution Strategy (i.e., $\beta$, $\gamma$, $\delta$, $\nu$) can be easily set by following state-of-the-art guidelines [29][30][31]. Moreover, it must be pointed out that the proposed *CS*-based *Pattern-Matching Contiguous-Clustering* approach directly works in the complex domain of the excitations, without requiring - as usually done in many *CS* methods for array synthesis [27] - a reformulation in a real-valued problem. This enables a more efficient and physically-admissible (i.e., equal support/element-locations for the real and the imaginary part of the excitation coefficients) setting of the *DoF*s.

## IV. Numerical Assessment

This section is aimed at assessing the features and the potentialities of the proposed clustering strategy when handling various array apertures and target patterns with different sidelobe levels ($SLL$s) also throughout some comparisons with competitive state-of-the-art subarraying methods. Towards this end and besides the graphical representations of the synthesized layouts and the corresponding radiated patterns, the *TV-CS* solutions will be evaluated in terms of the *pattern matching index*

Table I
*Comparative Assessment* ($N = 128$, *Taylor Pattern* [$SLL = -50$ dB]) - PERFORMANCE INDEXES.

| | *Pattern Indexes* | | | *Excitations Indexes* | | *Computational Index* |
|---|---|---|---|---|---|---|
| | $\xi$ $\left[\times 10^{-3}\right]$ | $D_{\max}$ [dB] | $SLL$ [dB] | $\chi$ $\left[\times 10^{-1}\right]$ | $DRR$ [dB] | $\Delta t$ [s] |
| ***Ref.*** | – | 19.58 | $-50.0$ | - | 25.14 | – |
| **[21]** | 6.44 | 19.89 | $-35.9$ | 1.17 | 22.14 | – |
| **[22]** | 3.80 | 19.70 | $-36.9$ | 1.17 | 20.44 | $1.59 \times 10^{2}$ |
| ***TV-CS*** | 2.76 | 19.66 | $-34.9$ | 1.17 | 19.80 | $7.30 \times 10^{-1}$ |
| ***TV-CS*** | 3.96 | 19.59 | $-30.7$ | 1.01 | 19.30 | $9.41 \times 10^{-1}$ |

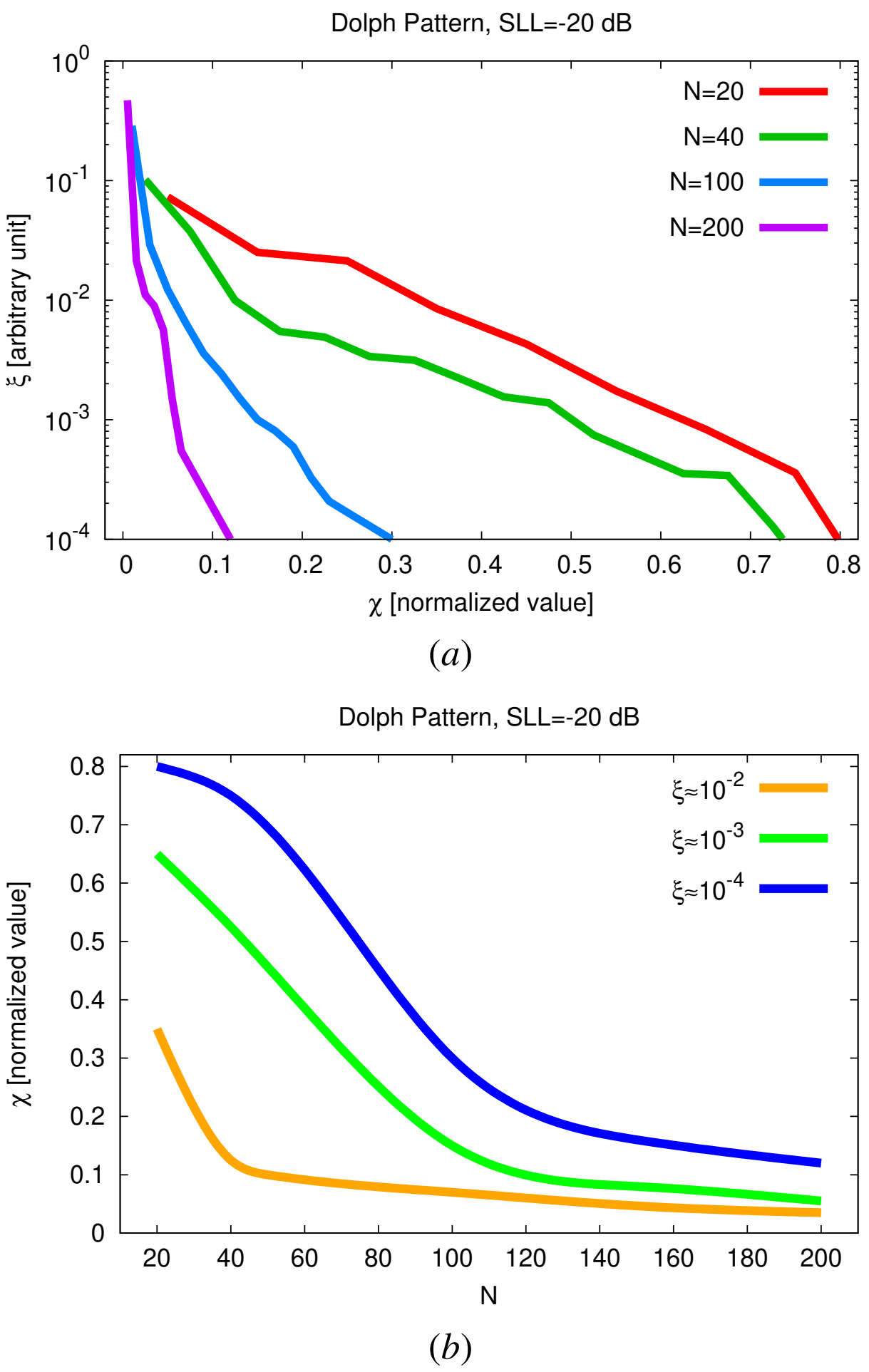

Figure 4. *Numerical Validation* (*Dolph Pattern* [$SLL = -20$ *dB*]) - Behaviour of (*a*) $\xi$ versus $\chi$ when $N \in (20, 200)$, and (*b*) $\chi$ vs. $N$ when $\xi \in \left\{10^{-4}, 10^{-3}, 10^{-2}\right\}$.

$$\xi = \frac{\int_0^{\pi}\left|\widetilde{f}(\theta) - f(\theta)\right|^2 \sin(\theta)\, d\theta}{\int_0^{\pi}\left|\widetilde{f}(\theta)\right|^2 \sin(\theta)\, d\theta} \tag{18}$$

versus the clustering factor $\chi \triangleq \frac{Q}{N}$, $Q$ being the number of control points in the array layout (i.e., the number of clusters) given by $Q \triangleq \|\nabla \widehat{\mathbf{w}}\|_1 + 1$ [Fig. 1(*c*)]. As for the *TV-CS* control parameters, they have been set according to the state-of-the-art guidelines in [29]: $\delta = 10^{-3}$ and $\nu = 10^{-5}$.

The first benchmark problem is concerned with the design of a $N = 100$ clustered arrangement of half-wavelength spaced isotropic radiators ($L = 49.5\lambda$ being the total aperture of the array) matching a Dolph pattern with $SLL = -20$ dB [Fig. 2(*a*)], $\widetilde{f}(\theta)$, sampled at $M = 67$ angular locations $\theta_m = \arcsin\left[\left(\frac{m-1}{M-1}\right)\right]$, $m = 1, ..., M$ [Fig. 2(*a*)] [14][40]. The performance of the *TV-CS* technique for different configurations of the control parameters ($\beta \in \left[10^{-10}, 10^{10}\right]$ and $\gamma \in \left[10^{-10}, 10^{10}\right]$ [29]) are resumed in Fig. 2(*b*) in terms of the matching error values $\xi$ versus the clustering factor $\chi$. The representative points of the *TV-CS* solutions in the $(\xi, \chi)$-plane and the resulting Pareto front indicate that, as expected, the accuracy of the matching with the reference pattern enhances as the complexity of the feed network increases [e.g., $\xi\rfloor_{\chi=0.05} \approx 1.22 \times 10^{-2}$ vs. $\xi\rfloor_{\chi=0.15} \approx 1.00 \times 10^{-3}$ - Fig. 2(*b*)]. Such a behaviour can be also visually verified by comparing the pattern of the reference *FP* array, $\widetilde{f}(\theta)$, with those of the *TV-CS* clustered solutions with $\chi = 0.05$, $f(\theta)\rfloor_{\chi=0.05}$, and $\chi = 0.15$, $f(\theta)\rfloor_{\chi=0.15}$ [Fig. 3(*a*)] corresponding to the representative points circled in Fig. 2(*b*). More specifically, the pattern matching improves especially in the sidelobe region close to the mainlobe [see the inset of Fig. 3(*a*)]. It is also worthwhile to notice that, even though the synthesis problem is formulated as a *pattern matching* one, greater $Q$ values ($\Rightarrow \chi \uparrow$) implicitly yield better *"staircase"* approximations of the reference excitations in the $\mathbf{w}$-excitation domain as shown in Fig. 3(*b*) where the vector $\mathbf{w}\rfloor_{\chi=0.05}$ and $\mathbf{w}\rfloor_{\chi=0.15}$ are represented together with the reference *FP* excitations.

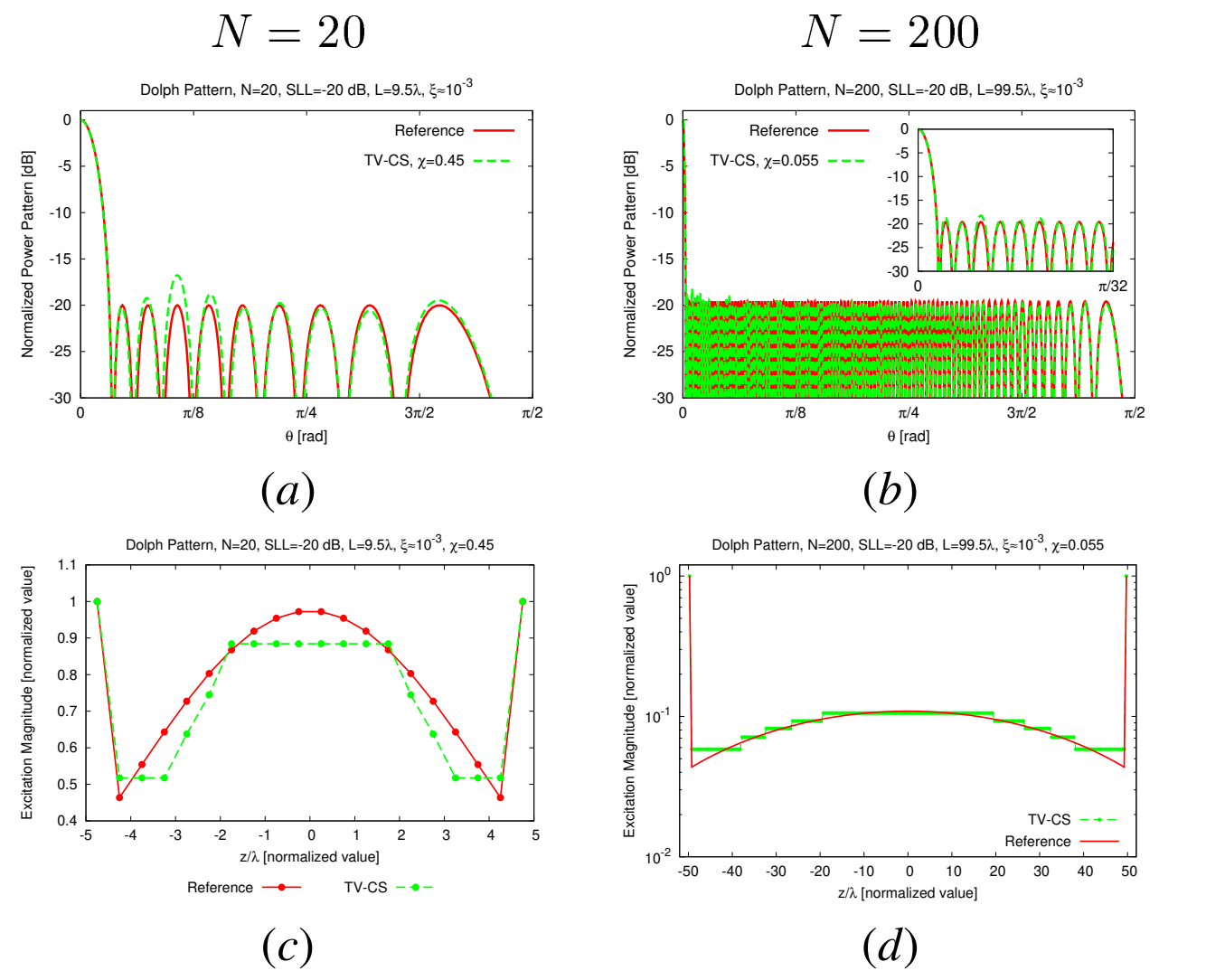

Figure 5. *Numerical Validation* (*Dolph Pattern* [$SLL = -20$ *dB*], $\xi \approx 10^{-3}$) - Reference and *TV-CS* patterns (*a*)(*b*) and corresponding excitations (*c*)(*d*) when (*a*)(*c*) $N = 20$ and (*b*)(*d*) $N = 200$.

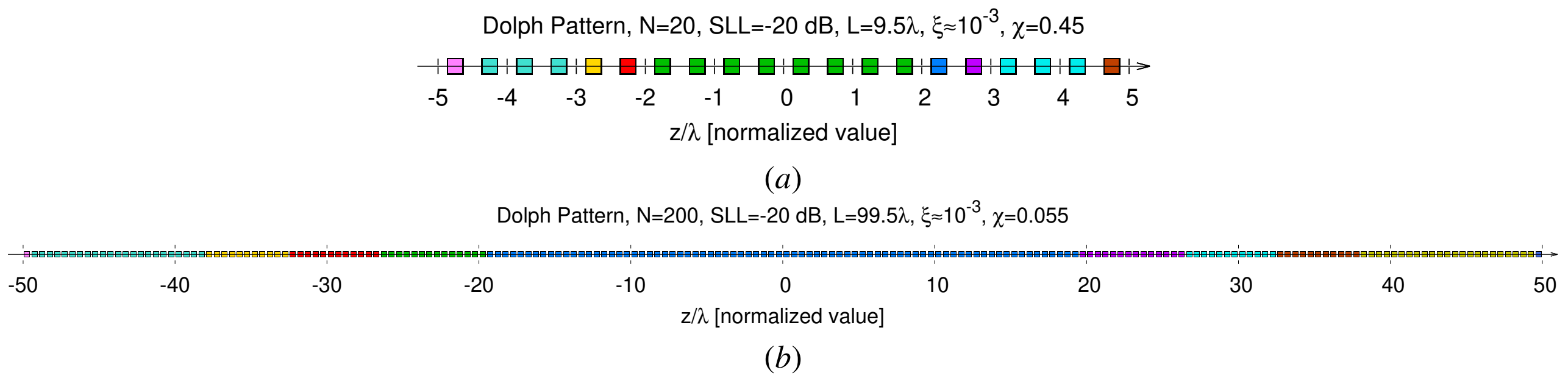

Figure 6. *Numerical Validation* (*Dolph Pattern* [$SLL = -20$ *dB*], $\xi \approx 10^{-3}$) - *TV-CS* clustering schemes when ($a$) $N = 20$ and ($b$) $N = 200$.

The architectural schemes of the corresponding feed networks [$\chi = 0.05$ - Fig. 3($c$) vs. $\chi = 0.15$ - Fig. 3($d$)] show that, unlike [14], the *TV-CS* approach synthesizes physically-contiguous clusters [Figs. 3($c$)-3($d$)] without element overlaps as well as it avoids the designer to impose some constraints on the subarrays size/shapes to be used. This is a non-negligible and non-trivial advantage since the synthesis process can now fully explore the solution space of the design *DoF*s without neglecting any admissible clusters arrangement. These considerations are confirmed by the results when changing the array size by varying the $N$ value (Fig. 4), while keeping as reference a Dolph pattern with $SLL = -20$ dB. Indeed, the plots of the *TV-CS* Pareto fronts in the $(\xi, \chi)$-plane [Fig. 4($a$)] show that, regardless of the array size, an increase of $\chi$ (i.e., a greater number of clusters $Q$) entails a better pattern matching [e.g., $\left.\frac{\xi\rfloor_{\chi=0.1}}{\xi\rfloor_{\chi=0.8}}\right\rfloor_{N=20} \approx 4.52 \times 10^2$ - Fig. 4($a$)]. The analysis also indicates that such an improvement is more significant for wider apertures, the Pareto front slopes being steeper for greater $N$ values [e.g., $\left.\frac{\chi\rfloor_{N=200}}{\chi\rfloor_{N=20}}\right\rfloor_{\xi=10^{-3}} \approx 9.1 \times 10^{-2}$ - Fig. 4($a$)]. This means that the wider the layouts are, the more the feed network can be simplified (i.e., $\chi$ reduces) for a given accuracy in matching the reference pattern. To further and better assess such a feature, the plots of $\chi$ versus the aperture size, $N$, for different $\xi$ values (i.e., $\xi \in \left\{10^{-4}, 10^{-3}, 10^{-2}\right\}$) are reported in Fig. 4($b$). As it can be observed, an improvement of the accuracy from $\xi = 10^{-2}$ down to $\xi = 10^{-4}$ needs increasing the clustering value from $\chi = 0.35$ up to $\chi = 0.8$ when $N = 20$ ($\Delta\chi = 0.45$), while it is enough moving from $\chi = 0.035$ up to $\chi = 0.12$ when $N = 200$ ($\Delta\chi = 0.085$). This is visually confirmed in Fig. 5 by the plots of the patterns [Figs. 5($a$)-5($b$)] radiated by the synthesized excitations [Fig. 5($c$)-5($d$)] when requiring a pattern matching of the order of $\xi \approx 10^{-3}$ and considering a smaller array with $N = 20$ elements and a larger one $N = 200$, respectively. More in detail, it turns out that the same accuracy [$\xi \approx 10^{-3}$ - Fig. 5($a$) vs. Fig. 5($b$)] corresponds to very different clustering factors [$\chi = 0.45$ - Fig. 5($c$) vs. $\chi = 0.055$ - Fig. 5($d$)] as well as complexity of the architecture of the associated feed networks [$N = 20$ - Fig. 6($a$) vs. $N = 200$ - Fig. 6($b$)].

The next numerical experiment is devoted to check the matching performance of the *TV-CS* clustering when varying the $SLL$ of the reference pattern. Accordingly, the layout with $N = 100$ elements radiating a Dolph pattern has been used as benchmark, while the $SLL$ has been changed in the range $SLL \in [-40\,dB, -20\,dB]$. Figure 7 gives the plots of the achievable clustering factor $\chi$ versus the target $SLL$ for different values of the accuracy index $\xi$ [Fig. 7($a$)]. It can be inferred that ($i$) to keep the same pattern matching, the complexity of the feed network has to be increased if more challenging $SLL$ requirements are at hand [e.g., $\chi\rfloor_{SLL=-20\,dB}^{\xi=10^{-4}} \approx 0.25$ vs. $\chi\rfloor_{SLL=-40\,dB}^{\xi=10^{-4}} \approx 0.75$ - Fig. 7($a$)], ($ii$) regardless of the target $SLL$, more clusters (i.e., greater values of $\chi$) are necessary to synthesize patterns closer to the reference one [e.g., $\left.\frac{\chi\rfloor_{\xi=10^{-4}}}{\chi\rfloor_{\xi=10^{-2}}}\right\rfloor_{SLL=-30\,dB} \approx 2.92$ - Fig. 7($a$)], ($iii$) a lower layout complexity is required for smaller values of the reference $SLL$ [e.g., $\left.\frac{\chi\rfloor_{\xi=10^{-4}}}{\chi\rfloor_{\xi=10^{-2}}}\right\rfloor_{SLL=-40\,dB} \approx 2.77$ vs. $\left.\frac{\chi\rfloor_{\xi=10^{-4}}}{\chi\rfloor_{\xi=10^{-2}}}\right\rfloor_{SLL=-20\,dB} \approx 5.0$ - Fig. 7($a$)]. For illustrative purposes, the plots of two representative *TV-CS* layouts are reported [Fig. 7($b$) and Fig, 7($d$) - $SLL = -30$ dB; Fig. 7($c$) and Fig. 7($e$) - $SLL = -40$ dB], as well. Analogously to the previous analysis on the array dimension, it turns out that lowering $\xi$ usually yields a less accurate sidelobe approximation in the angular region near the mainlobe [see the insets in Figs. 7($b$)-7($c$)] as well as, although exploiting a *pattern-matching* formulation and not an *excitation-matching* one, a worse approximation of the reference excitations [Figs. 7($d$)-7($e$)].

The dependence of the matching accuracy of the *TV-CS* method on the pattern shape has been assessed next by still considering a $N = 100$-elements array, but now radiating either a Taylor pattern with $SLL = -30$ dB [Fig. 8($c$)] or a "flat-top" pattern [Fig. 8($d$)]. The plots the *TV-CS* Pareto fronts in the $(\xi, \chi)$-plane [Figs. 8($a$)-8($b$)] indicate that the *TV-CS* qualitatively performs as in the previous test cases [e.g., Figs. 8($a$)-8($b$) vs. Fig. 4($a$)], but also they point out that the clustering ratio for a desired matching accuracy depends on the pattern shape and not only on the $SLL$ value [e.g., ($SLL - 30$ dB) $\Rightarrow$ $\chi\rfloor_{Taylor}^{\xi\approx10^{-3}} \approx 0.37$ - Fig. 8($a$) vs. $\chi\rfloor_{Dolph}^{\xi\approx10^{-3}} \approx 0.49$ - Fig. 7($a$)]. Moreover, it turns out that faithfully matching a shaped (e.g., flat-top) pattern usually needs more complex feed networks than matching a pencil-beam one with the same $SLL$, for instance $\chi\rfloor_{Flat-Top}^{\xi\approx10^{-3}} \approx 0.42$ [Fig. 8($a$)] vs. $\chi\rfloor_{Dolph}^{\xi\approx10^{-3}} \approx 0.15$ [Fig. 2($a$)]. For completeness, Figure 8 also provides the plots of the patterns [Figs. 8($c$)-8($d$)] and the excitations [Figs. 8($e$)-8($f$)] of the representative results yielding the value of $\xi \approx 10^{-3}$ for the pattern matching.

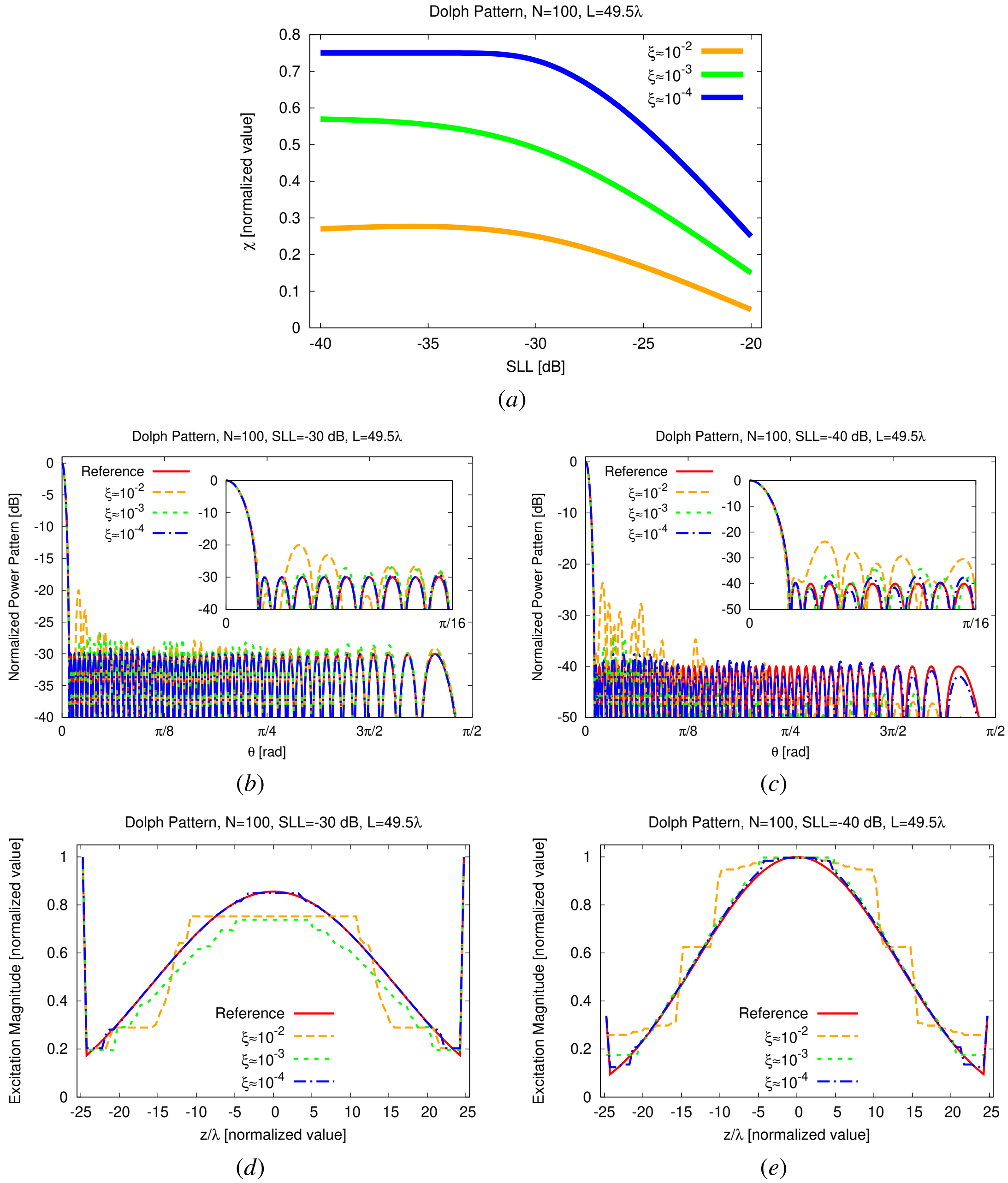

Figure 7. *Numerical Validation* ($N = 100$, *Dolph Pattern*) - Behaviour of $\xi$ versus $SLL$ ($SLL \in [-40, -20]$ *dB*) when $\xi \in \{10^{-4}, 10^{-3}, 10^{-2}\}$ $(a)$. Reference and *TV-CS* patterns $(b)(c)$ and the corresponding excitations $(d)(e)$ when $(b)(d)$ $SLL = -30$ dB and $(c)(e)$ $SLL = -40$ dB.

Once again, the above outcomes on both the *implicit* excitation matching capabilities of the *TV-CS* [Figs. 8($e$)-8($f$)] and its effectiveness in simplifying the feed network hold true even when very low sidelobes have to be carefully approximated [Figs. 8($c$)-8($d$)].

A comparative assessment of the *TV-CS* strategy with some state-of-the-art clustering methods, which handle spatially-contiguous subarrays without constraints on the subarray lengths and/or shapes [21][22] such as the proposed approach, is provided next. Towards this end, a $N = 128$ elements layout radiating a $SLL = -50$ [dB] Taylor pattern is taken into account (Fig. 9). With reference to the $(\xi, \chi)$-plane, let us compare the representative points of the clustered solutions obtained in [21] and [22] with those from the *TV-CS* synthesis. As for the pattern matching, it turns out that, setting a clustering index value $\chi$ (i.e., for an assigned architectural complexity of the feeding network), the *TV-CS* better matches the reference pattern thanks to its pattern-matching formulation. For instance, $\left.\frac{\xi\rfloor_{[Haupt\ 2007]}}{\xi\rfloor_{TV-CS}}\right|_{\chi=1.17\times10^{-1}} \approx 262\%$ and $\left.\frac{\xi\rfloor_{[Manica\ 2009]}}{\xi\rfloor_{TV-CS}}\right|_{\chi=1.17\times10^{-1}} \approx 155\%$ when $\chi = 1.17 \times 10^{-1}$ (Fig. 9 - Tab. I). Moreover, besides the value of the index $\xi$, which is the key and also a global indicator of the fitting with the reference, the effectiveness of the *TV-CS* in matching the target pattern is also pointed out by the values of the standard pattern indexes such as the directivity, $D_{max}$ ($\frac{D_{\max}\rfloor_{TV-CS} - D_{\max}\rfloor_{Ref.}}{D_{\max}\rfloor_{Ref.}} \approx 4.0 \times 10^{-3}$ vs. $\frac{D_{\max}\rfloor_{[Haupt\,2007]} - D_{\max}\rfloor_{Ref.}}{D_{\max}\rfloor_{Ref.}} \approx 1.5 \times 10^{-2}$ and $\frac{D_{\max}\rfloor_{[Manica\,2009]} - D_{\max}\rfloor_{Ref.}}{D_{\max}\rfloor_{Ref.}} \approx 6.0 \times 10^{-3}$ - Tab. I), while the *SLL* slightly worsens $[SLL\rfloor_{TV-CS} = -34.9$ [dB] vs. $SLL\rfloor_{[Haupt\,2007]} = -35.9$ [dB] and $SLL\rfloor_{[Manica\,2009]} = -36.9$ [dB] - Fig. 10$(a)$]. Concerning the complexity of the feeding network, the solutions found by the *TV-CS* method present more simple layouts than the state-of-the-art ones with less clusters. For examples, choosing the same pattern matching guarantee by the architecture in [22] (blue circle - Fig. 9) allows the *TV-CS* to synthesize a clustered arrangement (orange circle - Fig. 9) with a smaller $\chi$ (i.e., $\left.\frac{\xi\rfloor_{[Manica\,2009]}}{\chi\rfloor_{TV-CS}}\right|_{\chi=1.17\times10^{-1}} \approx 1.16$ - Tab. I). Moreover, not only the architecture of the feeding network is simplified, but also the implementation of the amplification chain is made easier since the dynamic range ratio ($DRR \triangleq \frac{\max_{n=1,..,N}|w_n|}{\min_{n=1,..,N}|w_n|}$) is also improved ($\left.\frac{DRR\rfloor_{Ref.} - DRR\rfloor_{TV-CS}}{DRR\rfloor_{Ref.}}\right|_{\chi=1.17\times10^{-1}} \approx 21.2\%$ vs. $\left.\frac{DRR\rfloor_{Ref.} - DRR\rfloor_{[Haupt\,2007]}}{DRR\rfloor_{Ref.}}\right|_{\chi=1.17\times10^{-1}} \approx 11.9\%$ and $\left.\frac{DRR\rfloor_{Ref.} - DRR\rfloor_{[Manica\,2009]}}{DRR\rfloor_{Ref.}}\right|_{\chi=1.17\times10^{-1}} \approx 18.6\%$ - Tab. I) as it can be observed in the plots of the corresponding excitations [Fig. 10$(b)$]. Such an outcome, which is confirmed regardless of the clustering factor (e.g., $\left.\frac{DRR\rfloor_{TV-CS} - DRR\rfloor_{Ref.}}{DRR\rfloor_{Ref.}}\right\rfloor_{\chi=1.01\times10^{-1}} \approx 23.2\%$ - Tab. I), is theoretically expected since minimizing $\|\nabla \mathbf{w}\|_1$ [see (2)] forces contiguous sub-arrays to have similar excitation magnitudes, thus the method implicitly looks for the excitations distribution that minimizes the $DRR$.

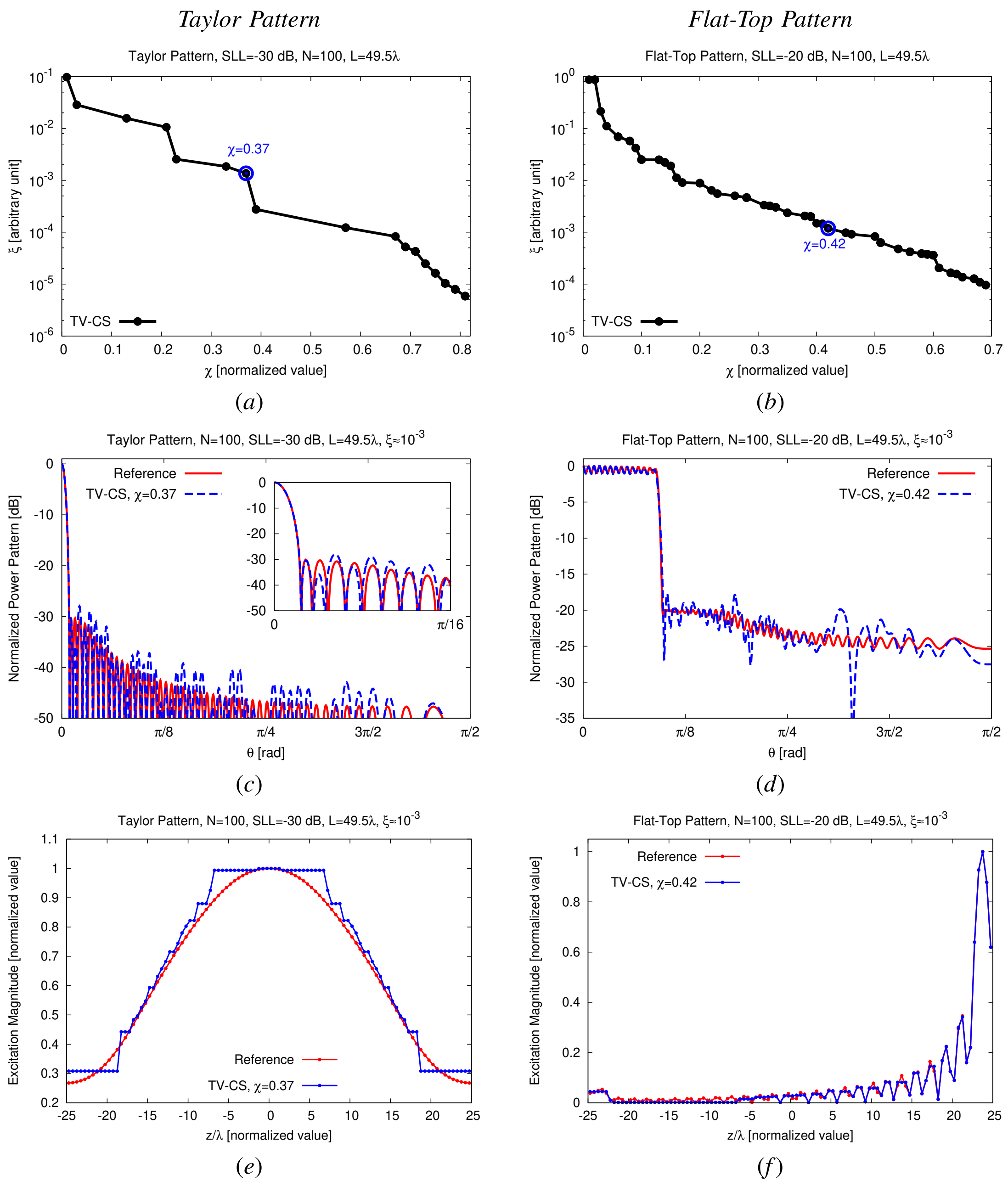

Figure 8. *Numerical Validation* ($N = 100$) - *TV-CS* Pareto fronts in the $(\xi, \chi)$-plane $(a)(b)$, reference and *TV-CS* patterns with $\xi \approx 10^{-3}$ $(c)(d)$, and the corresponding excitations $(e)(f)$ when matching $(a)(c)(e)$ a *Taylor* pattern [$SLL = -30$ dB] and $(b)(d)(f)$ a *flat-top* pattern [$SLL = -20$ dB].

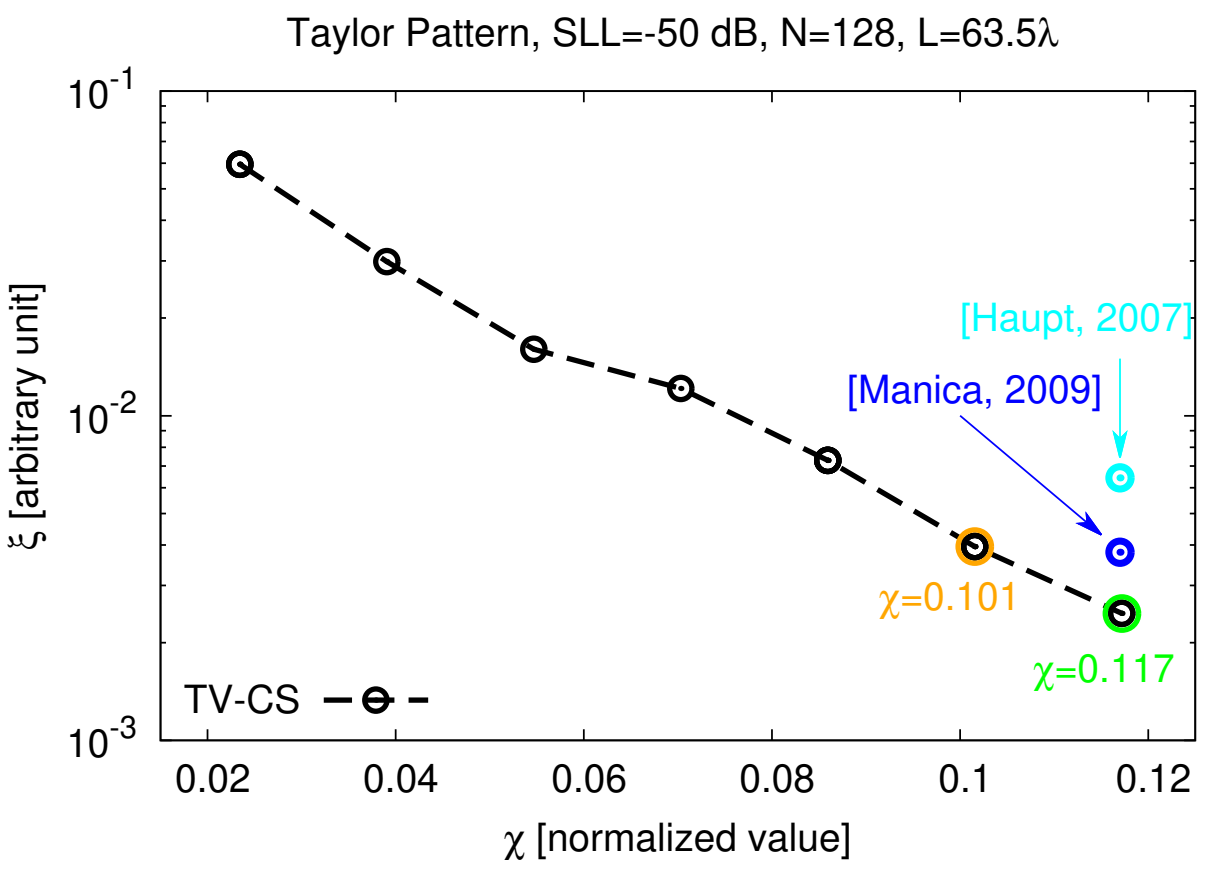

Figure 9. *Comparative Assessment* ($N = 128$, *Taylor Pattern* [$SLL = -50$ dB]) - Representative *TV-CS* and state-of-the-art solutions in the $(\xi, \chi)$-plane.

Taking a look to the computational indexes[2], as well, it is worth remarking that the exploitation of the customized deterministic *alternating direction* algorithm [30][31][32] as a *TV* sparseness-promoting solver allows one to reach an excellent numerical efficiency ($\Delta t\rfloor_{TV-CS} \leq 1$ [s] - Tab. I) also with respect to the local search techniques used by the other clustering approaches (e.g., $\frac{\Delta t\rfloor_{TV-CS} - \Delta t\rfloor_{[Manica\ 2009]}}{\Delta t\rfloor_{[Manica\ 2009]}} \approx -99.4\%$ - Tab. I). This confirms what foreseen in the *CS* literature [27] and already proven in other applications of the *TV-CS* formulation in electromagnetics [29].

The final numerical example is aimed at assessing the *TV-CS* reliability and its effectiveness when dealing with real array elements, that is, taking into account during the synthesis process of the impact of mutual coupling and non-ideal antenna patterns. More specifically, two linear arrangements of aperture-stacked patch antennas [41] with either $N = 20$ [Fig. 11(*a*)] or $N = 40$ [Fig. 11(*b*)] array elements have been considered and the reference pattern has been set to a Taylor one with $SLL = -20$ [dB]. To faithfully deal with such a non-ideal array, the radiating elements have been modeled according to [41] and the $n$-th ($n = 1, ..., N$) embedded element pattern $e_n(\theta)$ has been set to the *active* element one [2], once computed through a full-wave simulator, and then employed to compute $\mathcal{H}$ in (4). The plots of the radiated patterns from the reference and a representative trade-off set of *TV-CS* clustered excitations [Figs. 11(*c*)-11(*d*)] are shown in Fig. 12. As it can be quantitatively appreciated in Fig. 12 and better pointed out along the representative $\phi = 0$ pattern cut [Figs. 11(*e*)-11(*f*)], the *TV-CS* technique still reliably matches the reference pattern [i.e., $\xi \approx 3.71 \times 10^{-3}$ - Fig. 12(*a*) vs. Fig. 12(*c*); $\xi \approx 3.06 \times 10^{-3}$ - Fig. 12(*b*) vs. Fig. 12(*d*)] as well as the desired sidelobe shape also closely approximating both the directivity (i.e., $\left.\frac{D_{\max}\rfloor_{TV-CS} - D_{\max}\rfloor_{Ref.}}{D_{\max}\rfloor_{Ref.}}\right\rfloor_{N=20,\,40} \approx 3.9 \times 10^{-3}$ - Tab. II) and the $SLL$ reference values (Tab. II). On the other hand, once again and despite the presence of non-idealities and mutual-coupling effects in the definition of the radiation operator $\mathcal{H}$ (4), the proposed *pattern-matching* strategy implicitly yields an *excitation-matched* clustered layout [Figs. 11(*c*)-11(*d*)] with a non-negligible reduction of the feed network complexity (i.e., $\chi \leq 0.38$ and $DRR \leq 3.8$ [dB] - Tab. II). These outcomes further highlight the flexibility and the customizability of the proposed subarraying approach.

[2]For fair comparisons, all the syntheses have been performed with non-optimized MATLAB versions of the SW codes executed on a single-core CPU running at 2 GHz.

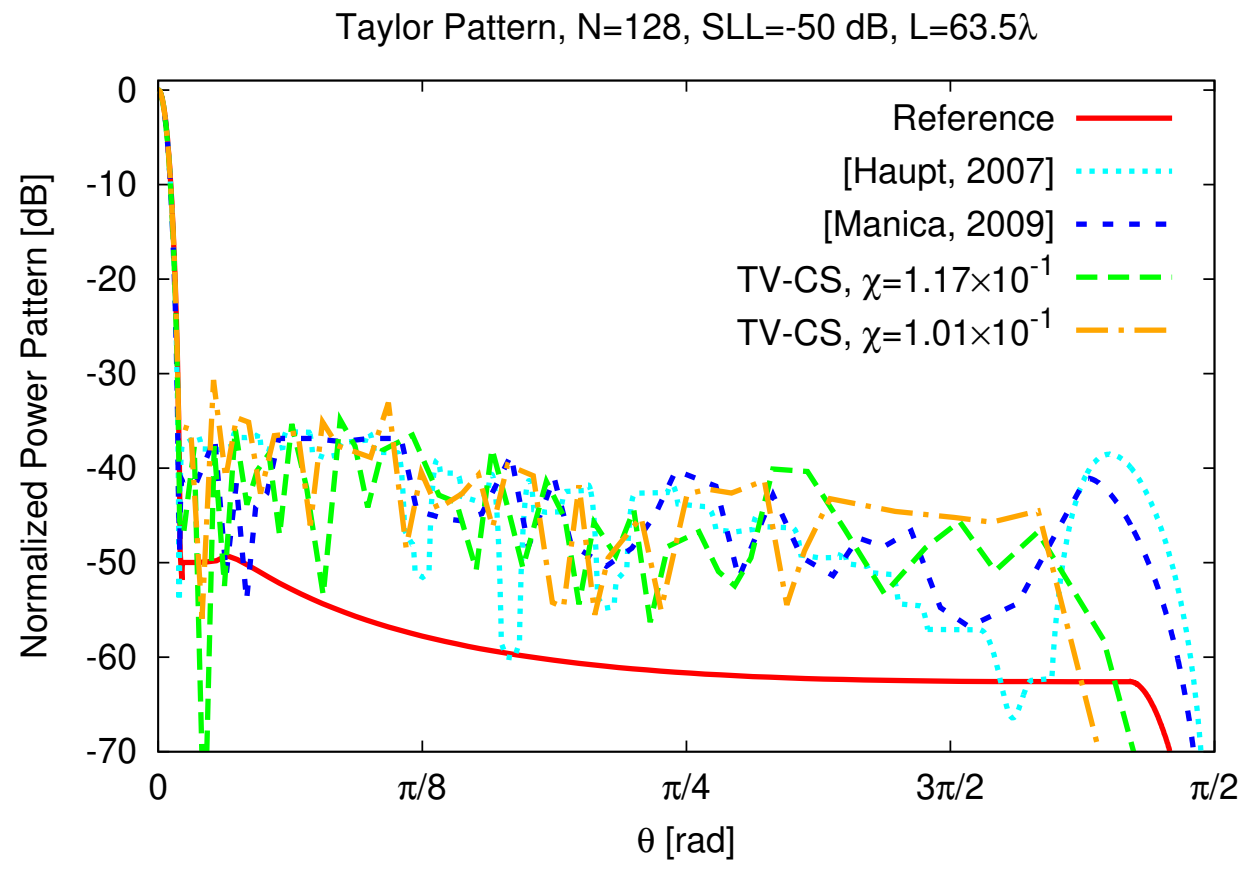

(*a*)

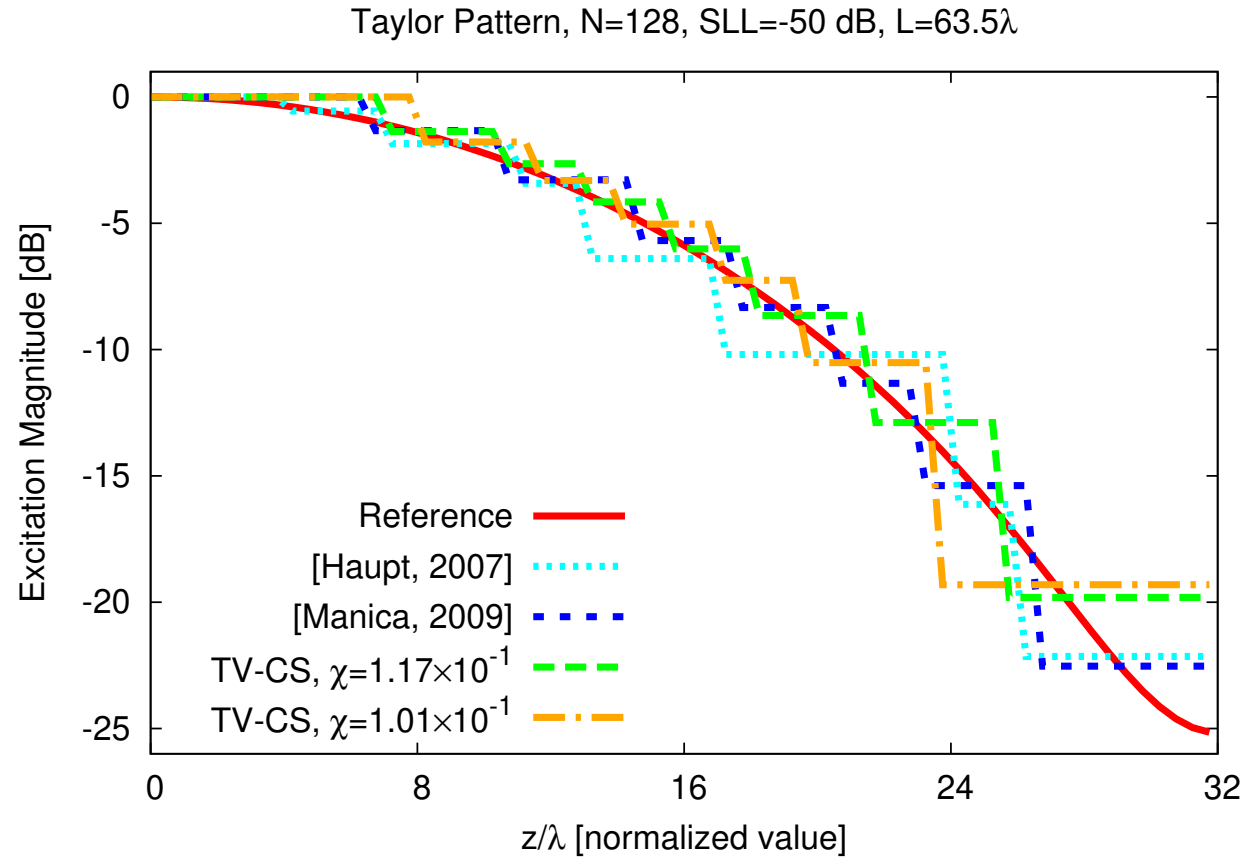

(*b*)

Figure 10. *Comparative Assessment* ($N = 128$, *Taylor Pattern* [$SLL = -50$ dB]) - Reference, state-of-the-art [21][22], and *TV-CS* patterns (*a*) and the corresponding excitations (*b*).

## V. Conclusions and Remarks

The design of contiguously-clustered linear arrays matching a user-defined reference pattern has been addressed with a new method leveraging on a *TV-CS* formulation of the subarraying problem. More specifically, the computation of the clustered feed network coefficients has been first recast as the maximization of the sparsity of the *gradient* of the array excitations distribution subject to matching the reference pattern. Then, the arising *TV-CS* problem has been solved by means of a customized version of a deterministic *alternating direction* algorithm originally proposed for video sensing applications [30][31][32]. A set of representative results has been presented to give some indications on the features and the potentialities of the proposed synthesis approach when dealing with both ideal and realistic antenna elements. Some comparisons with competitive state-of-the-art subarraying techniques have been carried out, as well.

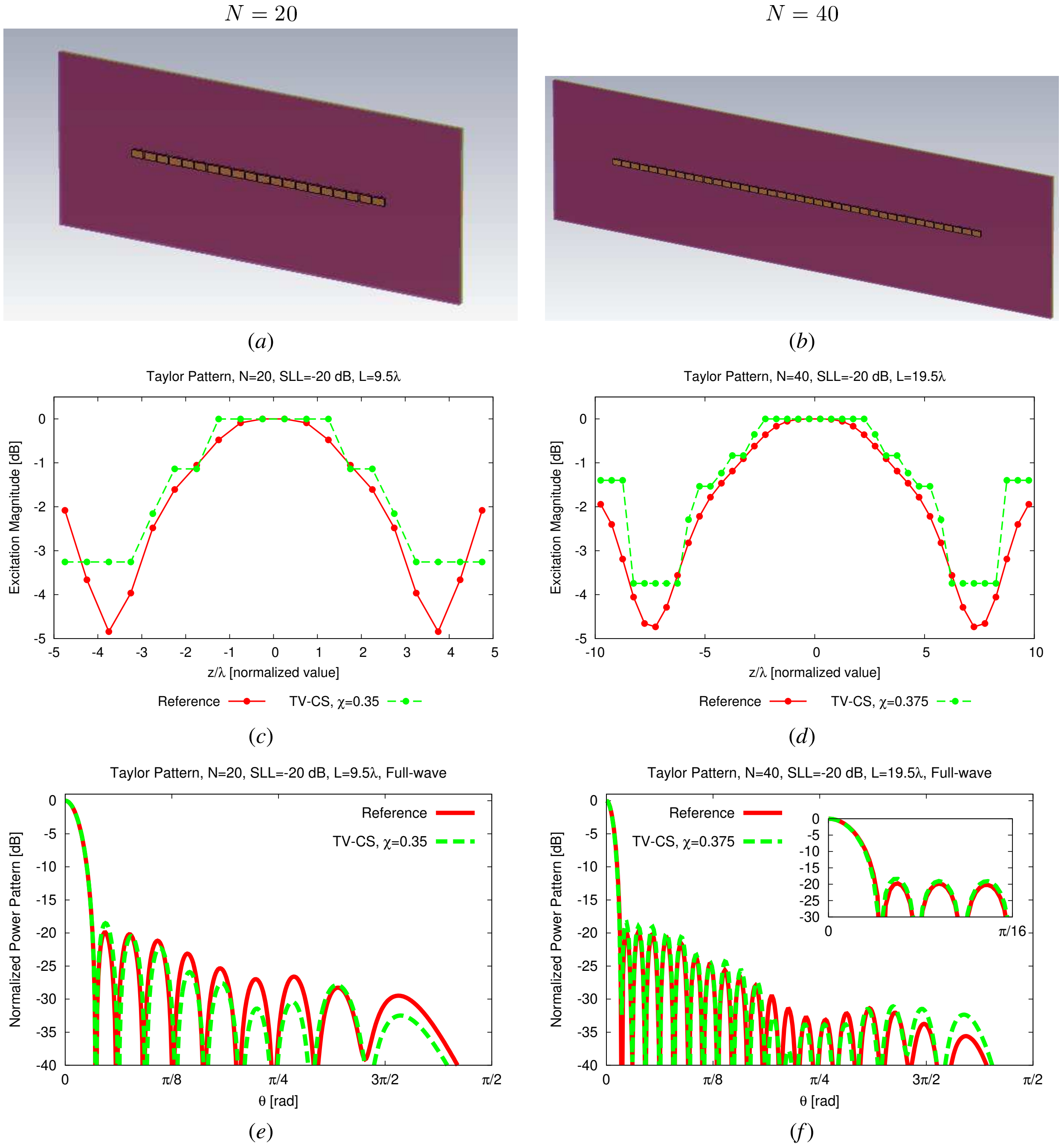

Figure 11. *Numerical Validation* (*Aperture-Stacked Patch Antenna Array* [41], *Taylor Pattern* [$SLL = -20$ dB]) - 3D array models (*a*)(*b*), *FP* reference and *TV-CS* clustered excitations (*c*)(*d*) and the associated full-wave simulated patterns (*e*)(*f*) when (*a*)(*c*)(*e*) $N = 20$ and (*b*)(*d*)(*f*) $N = 40$.

Table II
*Numerical Validation* (*Aperture-Stacked Patch Linear Array* [41], *Taylor Pattern* [$SLL = -20$ dB]) - PERFORMANCE INDEXES.

| | | ***Pattern Indexes*** | | | ***Excitations Indexes*** | |
|---|---|---|---|---|---|---|
| | $N$ | $\xi$ $[\times 10^{-3}]$ | $D_{\max}$ [dB] | $SLL$ [dB] | $\chi$ $[\times 10^{-1}]$ | $DRR$ [dB] |
| ***Ref.*** | 20 | – | 18.02 | $-19.92$ | – | 4.84 |
| **TV-CS** | 20 | 3.71 | 18.09 | $-18.51$ | 0.35 | 3.26 |
| ***Ref.*** | 40 | – | 20.45 | $-19.76$ | – | 4.74 |
| **TV-CS** | 40 | 3.06 | 20.53 | $-18.37$ | 0.375 | 3.74 |

The numerical assessment has shown that

- the *TV-CS* technique is able to derive a set of Pareto-optimal solutions in the $(\xi, \chi)$-plane for each user-defined

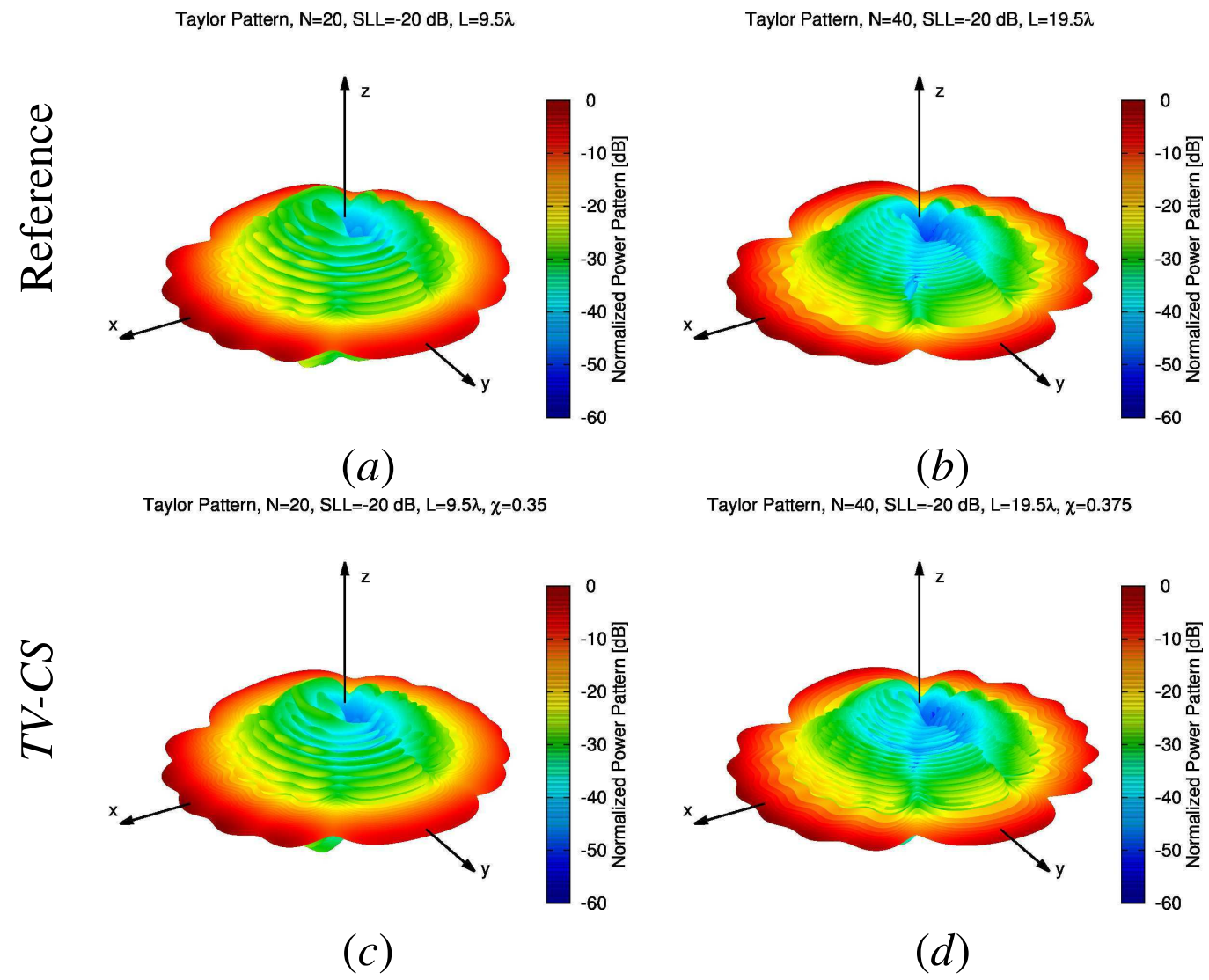

Figure 12. *Numerical Validation* (*Aperture-Stacked Patch Linear Array* [41], *Taylor Pattern* [$SLL = -20$ dB]) - 3D full-wave simulated pattern of the reference *FP* ($a$)($b$) and *TV-CS* clustered ($c$)($d$) layouts when ($a$)($c$) $N = 20$ and ($b$)($d$) $N = 40$.

target/reference pattern. Each representative point of such a Pareto front can be selected by the designer depending on the desired trade-off between matching accuracy, $\xi$, and feed-network complexity, $\chi$ [e.g., Fig. 2($b$), Fig. 4, Fig. 7($a$), Figs. 8($a$)-8($b$), and Fig. 9];

- despite the *pattern-matching* formulation, the proposed technique implicitly forces a staircase approximation of the reference layout in the *excitation* domain [e.g., Fig. 3($b$)];
- as expected, the higher the accuracy in matching the reference pattern, $\xi$, the greater is the complexity of the architecture of the *TV-CS* feed network, $\chi$, as well as the corresponding number of required clusters, $Q$, for a fixed aperture (i.e., number of array elements, $N$);
- for a fixed value of the pattern matching, $\xi$, the complexity of the *TV-CS* clustered scheme, $\chi$, reduces proportionally to the array dimension (i.e., improved clustering factors arise for wider apertures [e.g., Fig. 5]) and inversely to the pattern shape/*SLL*;
- the *TV-CS* method positively compares with competitive state-of-the-art unconstrained sub-arraying approaches in terms of matching accuracy, $\chi$ reduction, $DRR$ mitigation, and computational efficiency (e.g., Fig. 9 and Tab. I);
- the proposed technique is not restricted to ideal arrays, but it can be profitably exploited also when realistic radiating elements are at hand (e.g., Fig. 11) to yield non-negligible feed network simplifications with respect to the corresponding *FP* architectures (e.g., $\chi \leq 0.38$ and $DRR \leq 3.8$ [dB] - Tab. II), while reaching a satisfactory matching accuracy ($\xi \leq 3.71 \times 10^{-3}$ - Tab. II).

Moreover, the main and most innovative methodological advancements of this paper with respect to the state-of-the-art include: ($i$) the formulation of the clustering problem as a reference *pattern matching* and a sparseness-promoting one that operates in the *gradient excitation-domain*; ($ii$) the derivation of a *CS* sub-arraying algorithm that naturally avoids the overlaps among contiguous clusters and it does not require the *a-priori* definition of a *cluster dictionary*; ($iii$) the extension and the customization of the deterministic *alternating direction* algorithm in [30][31][32] to array synthesis, and ($iv$) the formulation of a *CS*-based clustering method working in the complex $\mathbf{w}$ domain, unlike most *CS* algorithms [27][37]. Future works, beyond the scope of the present paper, will be aimed at extending the proposed strategy to more complex geometries (e.g., planar/conformal) as well as to address the synthesis of multi-function arrays.

## Acknowledgements

A. Massa wishes to thank E. Vico for her never-ending support and help.

## References

[1] R. C. Hansen, *Phased Array Antennas*, 2nd Ed. Hoboken, NJ: Wiley, 2008.

[2] R. J. Mailloux, *Phased Array Antenna Handbook.* Boston, MA: Artech House, 2005.

[3] R. J. Mailloux, "Phased array theory and technology," *Proc. IEEE*, vol. 70, no. 3, pp. 246-291, Mar. 1982.

[4] J. E. Stailey and K. D. Hondl, "Multifunction phased array radar for aircraft and weather surveillance," *Proc. IEEE*, vol. 104, no. 3, pp. 649-659, Mar. 2016.

[5] M. Ardeshir Tanha, P. V. Brennan, M. Ash, A. Kohler and J. McElwaine, "Overlapped phased array antenna for avalanche radar," *IEEE Trans. Antennas Propag.*, vol. 65, no. 8, pp. 4017-4026, Aug. 2017.

[6] N. Shinohara, "Beam control technologies with a high-efficiency phased array for microwave power transmission in Japan," *Proc. IEEE*, vol. 101, no. 6, pp. 1448-1463, Jun. 2013.

[7] G. Oliveri, L. Poli, and A. Massa, "Maximum efficiency beam synthesis of radiating planar arrays for wireless power transmission," *IEEE Trans. Antennas Propag.*, pp. 2490-2499, vol. 61, no. 5, May 2013.

[8] A. Massa, G. Oliveri, F. Viani, and P. Rocca, "Array designs for long-distance wireless power transmission - State-of-the-art and innovative solutions," *Proc. IEEE*, vol. 101, no. 6, pp. 1464-1481, Jun. 2013.

[9] A. Jacomb-Hood and E. Lier, "Multibeam active phased arrays for communications satellites," *IEEE Microw. Mag.*, vol. 1, no. 4, pp. 40-47, Dec. 2000.

[10] R. Telikepalli, P. C. Strickland, K. R. McKay, and J. S. Wight, "Wide band microstrip phased array for mobile satellite communications," *IEEE Trans. Microw. Theory Techn.*, vol. 43, no. 7, pp. 1758-1763, Jul. 1995.

[11] G. Han, B. Du, W. Wu, and B. Yang, "A novel hybrid phased array antenna for satellite communication on-the-move in Ku-band," *IEEE Trans. Antennas Propag.*, vol. 63, no. 4, pp. 1375-1383, Apr. 2015.

[12] G. Oliveri, G. Gottardi, F. Robol, A. Polo, L. Poli, M. Salucci, M. Chuan, C. Massagrande, P. Vinetti, M. Mattivi, R. Lombardi, and A. Massa, "Co-design of unconventional array architectures and antenna elements for 5G base stations," *IEEE Trans. Antennas Propag.*, vol. 65, no. 12, pp. 6752- 6766, Dec. 2017.

[13] P. Rocca, G. Oliveri, R. J. Mailloux, and A. Massa, "Unconventional phased array architectures and design methodologies - A review," *Proc. IEEE*, vol. 104, no. 3, pp. 544-560, March 2016.

[14] G. Oliveri, M. Salucci, and A. Massa, "Synthesis of modular contiguously clustered linear arrays through a sparseness-regularized solver," *IEEE Trans. Antennas Propag.*, vol. 64, no. 10, pp. 4277-4287, Oct. 2016.

[15] R. L. Haupt, "Reducing grating lobes due to subarray amplitude tapering," *IEEE Trans. Antennas Propag.*, vol. 33, no. 8, pp. 846-850, Aug. 1985.

[16] A. P. Goffer, M. Kam, and P. R. Herczfeld, "Design of phased arrays in terms of random subarrays," *IEEE Trans. Antennas Propag.*, vol. 42, no. 6, pp. 820-826, Jun. 1994.

[17] D. Petrolati, P. Angeletti, and G. Toso, "A lossless beam-forming network for linear arrays based on overlapped subarrays," *IEEE Trans. Antennas Propag.*, vol. 62, no. 4, pp. 1769 -1778, Apr. 2014.

[18] D. A. McNamara, "Synthesis of sub-arrayed monopulse linear arrays through matching of independently optimum sum and difference excitations," *IEE Proc. H Microwave Antennas Propag.*, vol. 135, no. 5, pp. 371-374, 1988.

[19] L. Manica, P. Rocca, A. Martini, and A. Massa, "An innovative approach based on a tree-searching algorithm for the optimal matching of independently optimum sum and difference excitations," *IEEE Trans. Antennas Propag.*, vol. 56, no. 1, pp. 58-66, Jan. 2008.

[20] P. Rocca, L. Manica, R. Azaro, and A. Massa, "A hybrid approach for the synthesis of sub-arrayed monopulse linear arrays," *IEEE Trans. Antennas Propag.*, vol. 57, no. 1, pp. 280-283, Jan. 2009.

[21] R. L. Haupt, "Optimized weighting of uniform subarrays of unequal size," *IEEE Trans. Antennas Propag.*, vol. 55, no. 4, pp. 1207-1210, Apr. 2007.

[22] L. Manica, P. Rocca, and A. Massa, "Design of subarrayed linear and planar array antennas with SLL control based on an excitation matching approach," *IEEE Trans. Antennas Propag.*, vol. 57, no. 6, pp. 1684-1691, Jun. 2009.

[23] P. Rocca, R. J. Mailloux, and G. Toso, "GA-based optimization of irregular sub-array layouts for wideband phased arrays design," *IEEE Antennas Wireless Propag. Lett.*, vol. 14, pp. 131-134, 2015.

[24] S. K. Goudos, K. A. Gotsis, K. Siakavara, E. E. Vafiadis, and J. N. Sahalos, "A multi-objective approach to subarrayed linear antenna arrays design based on memetic differential evolution," *IEEE Trans. Antennas Propag.*, vol. 61, no. 6, pp. 3042-3052, Jun. 2013.

[25] P. Rocca, M. Benedetti, M. Donelli, D. Franceschini, and A. Massa, "Evolutionary optimization as applied to inverse problems," *Inverse Problems*, vol. 25, pp. 1-41, Dec. 2009.

[26] P. Rocca, G. Oliveri, and A. Massa, "Differential Evolution as applied to electromagnetics," *IEEE Antennas Propag. Mag.*, vol. 53, no. 1, pp. 38-49, Feb. 2011.

[27] A. Massa, P. Rocca, and G. Oliveri, "Compressive sensing in electromagnetics - A review," *IEEE Antennas Propag. Mag.*, vol. 57, no. 1, pp. 224-238, Feb. 2015.

[28] P. Shah, U. K. Khankhoje, and M. Moghaddam, "Inverse scattering using a joint $L1$-$L2$ norm-based regularization," *IEEE Trans. Antennas Propag.*, vol. 64, no. 4, pp. 1373-1384, Apr. 2016.

[29] G. Oliveri, N. Anselmi, and A. Massa, "Compressive sensing imaging of non-sparse 2D scatterers by a total-variation approach within the Born approximation," *IEEE Trans. Antennas Propag.*, vol. 62, no. 10, pp. 5157-5170, Oct. 2014.

[30] C. Li, *An Efficient Algorithm for Total Variation Regularization with Applications to the Single Pixel Camera and Compressive Sensing*, Master Thesis, Rice University, Sep. 2009.

[31] C. Li, *Compressive Sensing for 3D Data Processing Tasks: Applications, Models, and Algorithms*. PhD Thesis, Rice University, Apr. 2011.

[32] C. Li, H. Jiang, P. Wilford, Y. Zhang, and M. Scheutzow, "A new compressive video sensing framework for mobile broadcast," *IEEE Trans. Broadcast.*, vol. 59, no. 1, pp. 197-205, Mar. 2013.

[33] D. Goldfarb and W. Yin, "Second-order cone programming methods for total variation based image restoration," *SIAM J. Scient. Comput.*, vol. 27, no. 2, pp. 622-645, 2005.

[34] E. J. Candes and T. Tao, "Decoding by linear programming," *IEEE Trans. Inf. Theory*, vol. 51, no. 12, pp. 4203-4215, Dec. 2005.

[35] J. Bioucas-Dias and M. Figueiredo, "A new TwIST: two-step iterative thresholding algorithm for image restoration," *IEEE Trans. Imag. Process.*, vol. 16, no. 12, pp. 2992-3004, Dec. 2007.

[36] S. Becker, J. Robin, and E. Candes, *NESTA: a fast and accurate first-order method for sparse recovery*. Tech. Rep., California Institute of Technology, Apr. 2009.

[37] G. Oliveri, M. Carlin, and A. Massa, "Complex-weight sparse linear array synthesis by Bayesian Compressive Sampling", *IEEE Trans. Antennas Propag.*, vol. 60, no. 5, pp. 2309-2326, May 2012.

[38] G. Oliveri and A. Massa, "Bayesian compressive sampling for pattern synthesis with maximally sparse non-uniform linear arrays," *IEEE Trans. Antennas Propag.*, vol. 59, no. 2, pp. 467-481, Feb. 2011.

[39] S. Nash and A. Sofer, *Linear and Non-Linear Programming*. McGraw-Hill Higher Education: Burr Ridge, US, 1996.

[40] G. Oliveri, P. Rocca, and A. Massa, "Reliable diagnosis of large linear arrays - A bayesian compressive sensing approach", *IEEE Trans. Antennas Propag.*, vol. 60, no. 10, pp. 4627-4636, Oct. 2012.

[41] S. D. Targonski, R. B. Waterhouse, and D. M. Pozar, "Design of wide-band aperture-stacked patch microstrip antennas," *IEEE Trans. Antennas Propag.*, vol. 46, no. 9, pp. 1254-1251, Sep. 1998.